\documentclass[aps,showpacs,amsmath,amssymbd,dvips,showkeys,nofootinbib,preprintnumbers]{revtex4}

\usepackage{longtable}
\usepackage{supertabular}
\usepackage{float}
\usepackage{tikz}
\usetikzlibrary{arrows,shapes,positioning}
\usepackage{pgfplots}
\usetikzlibrary{arrows}
\usepackage{relsize}
\usepackage{lscape}
\usepackage{graphicx}
\usepackage{color}
\def \be  {\begin{equation}}
\def \ee  {\end{equation}}
\def \ee  {\end{equation}}
\def \bea {\begin{eqnarray}}
\def \eea {\end{eqnarray}}

\def \le   {\left}
\def \ri   {\right}

\newcommand{\nn}{\nonumber}

\newcommand{\dslash}{\ensuremath{\partial\hspace{-1.2ex} /}}
\newcommand{\bsig}{\ensuremath{\bar{\sigma}}}

\begin{document}

\preprint{ECTP-2016-12}
\preprint{WLCAPP-2016-12}
\hspace{0.05cm}
%\title{Chiral phase structure of meson states}

\title{Chiral phase structure and sixteen meson states in SU($3$) Polyakov linear-sigma model at finite temperature and chemical potential in strong magnetic field}

\author{Abdel Nasser Tawfik}
\email{atawfik@nu.edu.eg}
\affiliation{Nile University, Juhayna Square of 26th-July-Corridor, 12588 Giza, Egypt}
\affiliation{Frankfurt Institute for Advanced Studies, Ruth Moufang Str. 1, 60438 Frankfurt, Germany}
\affiliation{World Laboratory for Cosmology And Particle Physics (WLCAPP), 11571 Cairo, Egypt}

\author{Abdel Magied Diab}
\email{a.diab@eng.mti.edu.eg}
\affiliation{Egyptian Center for Theoretical Physics (ECTP), MTI University, 11571 Cairo, Egypt}
\affiliation{World Laboratory for Cosmology And Particle Physics (WLCAPP), 11571 Cairo, Egypt}

\author{M. T. Hussein}
\email{tarek@sci.cu.edu.eg}
\affiliation{Physics Department, Faculty of Science,  Cairo University, 12613 Giza, Egypt}

\date{\today}

\begin{abstract}
In characterizing the chiral phase-structure of pseudoscalars ($J^{pc}=0^{-+}$), scalars ($J^{pc}=0^{++}$), vectors ($J^{pc}=1^{--}$) and axial-vectors ($J^{pc}=1^{++}$) meson states and their dependence on temperature, chemical potential, and magnetic fields, we utilize SU($3$) Polyakov linear-sigma model (PLSM) in mean-field approximation. We first determine the chiral (non)strange quark condensates, $\sigma_l$ and $\sigma_s$ and the corresponding deconfinement order parameters, $\phi$ and $\phi^*$, respectively, in thermal and dense (finite chemical potential) medium and finite magnetic field. The temperature and the chemical potential characteristics of nonet meson states normalized to the lowest {\it bosonic} Matsubara frequencies are analyzed. We noticed that all normalized meson masses become temperature independent at different {\it critical} temperatures. We observe that the chiral and deconfinement phase transitions are shifted to lower {\it quasicritical} temperatures with increasing chemical potential and magnetic field. Thus, we conclude that the magnetic field seems to have almost the same effect as that of the chemical potential, especially on accelerating the phase transition, i.e. inverse magnetic catalysis. We also find that increasing chemical potential enhances the mass degeneracy of the various meson masses, while increasing the magnetic field seems to reduce the critical chemical potential, at which the chiral phase transition takes place. Our mass spectrum calculations agree well with the recent PDG compilations and PNJL, lattice QCD calculations, and QMD/UrQMD simulations.
\end{abstract}


\maketitle

\tableofcontents
\makeatletter
\let\toc@pre\relax
\let\toc@post\relax
\makeatother

%--------------------------------------------------------------------------------
%                                        Introductions
%--------------------------------------------------------------------------------
\section{Introduction \label{intro}}

Quantum Chromodynamics (QCD) predicts that the hadron-quark phase transition, where the hadronic matter goes through a partonic phase transition; a new-state-of-matter or colored quark-gluon plasma (QGP), takes place under extreme conditions of  density (large chemical potential\footnote{In cosmological context, low baryon density used to be related to low baryon chemical potential. But both quantities are apparently distinguishable, for instance, the phase diagrams $T$-$\mu$ and $T$-$\rho$ aren't the same \cite{Peter2017}.}) and temperature.  Accordingly,  at high temperature and/or chemical potential, the confined hadrons are conjectured to dissolve into free colored quarks and gluons. Various heavy-ion experiments aims at characterizing the properties of this new-state-of-matter, for instance, the STAR at the Relativistic Heavy-Ion Collider (RHIC) at BNL, the ALICE at the Large Hadron Collider (LHC) at CERN and the future CBM at the Facility for Antiproton and Ion Research (FAIR) at GSI and the future MPD at the Nuclotron based Ion Collider fAcility (NICA) at JINR.    

The explicit chiral-symmetry breaking in QCD is assumed to contribute to the masses of the elementary particles \cite{Svetitsky:1986, MeyerOrtmanns:1996, Rischke:2004}. In QCD-like models, which incorporate some features of QCD so that they become even partly able to draw an approximate picture about what the first-principle lattice QCD simulations can do, such as, the Polyakov linear-sigma model (PLSM) or the Polyakov quark-meson (PQM) and the Polyakov Nambu-Jona Lasinio (PNJL) models, some properties of the QCD meson sectors could be been studied, numerically. In the present paper, we study the chiral phase-structure of sixteen meson states in finite magnetic field and in dependence on temperature and chemical potential. In characterizing the dependences of various meson masses of pseudoscalars ($J^{pc}=0^{-+}$), scalars ($J^{pc}=0^{++}$), vectors ($J^{pc}=1^{--}$) and axial-vectors ($J^{pc}=1^{++}$) meson states on temperature, chemical potential and magnetic field, we utilize SU($3$) PLSM in the mean-field approximation.  The magnetic field is a significant ingredient to be taken consideration, as it in relativistic heavy-ion collisions likely can reach as much as $10^{19}~$Gauss. It is also conjectured that the deconfinement phase transition comes up with considerable contributions to the meson mass spectra. We shall discuss on the conditions under which every certain meson state shall dissolve into free {\it colored} quarks and gluons and which temperature and chemical potential would be able to modify the chiral phase-structure of each of the nonets meson states (sixteen).

Not only peripheral heavy-ion collisions generate enormous magnetic fields, but also central ones, as well. Due to oppositely directed relativistic motion of charges (spectators) especially in off-central collisions and/or local momentum imbalance of the participants, a huge short lived magnetic field can be created. At SPS, RHIC and LHC energies, for instance, the mean values of such magnetic field have been estimated as $\sim 0.1$, $\sim 1$, and $\sim 15\, m_{\pi}^2$, respectively \cite{Skokov:2009, Elec:MagnetA,Elec:MagnetB}, where $m_{\pi}^2 \simeq 10^{18}$ Gauss. It should be emphasized that such estimation have been calculated within the Ultrarelativistic Quantum Molecular Dynamics model (UrQMD) at various impact parameters, i.e. different centralities.

It is noteworthy highlighting the remarkable influence of the Polyakov-loop potentials on scalar and pseudoscalar meson sectors. With the in(ex)clusion of the Polyakov-loop potentials to the LSM Lagrangian, the chiral phase-structure of the nonet meson states at finite temperatures has been evaluated \cite{Schaefer:2009}. The same study was carried out with(out) axial anomaly term \cite{V. Tiwari:2009, V. Tiwari:2013}. Authors AT and AD published a systematic study for the thermal (temperature) and dense (chemical potential) influences on the chiral phase-structures of the nonet meson states (sixteen) with(out) axial anomaly term and also with in(ex)clusion of the Polyakov-loop potentials \cite{Tawfik:2014gga}. For the sake of completeness, we recall that at finite chemical potential SU($3$) NJL model \cite{NJL:2013} and SU($3$) PNJL model \cite{P. Costa:PNJLA,P. Costa:PNJLB,P. Costa:PNJLC} have been applied to characterize the nonet meson states, as well. 

The present work represents an extension of Ref. \cite{Tawfik:2014gga} to finite magnetic field. We present a systematic study of the influences of finite magnetic field, which is likely in high-energy collisions, and the possible modifications on the chiral phase-structure of various mesonic states (sixteen), including (pseudo) scalar and (axial) vector meson sectors, as functions of temperatures and chemical potential in presence of finite magnetic field. The temperature and chemical potential dependences of the chiral nonstrange  and strange quark condensates, $\sigma_l$ and $\sigma_s$ and the corresponding deconfinement order-parameters, $\phi$ and $\phi^*$, respectively, determining the quark-hadron phase transitions in finite magnetic field shall be characterized at finite magentic field. Then, we investigate the various nonet meson states. Last but not least, we estimate how the normalization of the these meson states relative to the lowest {\it bosonic} Matsubara frequencies behaves with temperature,  chemical potential and magnetic field. This allows us an approximate determination of the dissolving temperature and chemical potential for each meson state in varying magnetic field.

An extensive comparison between our calculations of the meson nonets and the latest compilation of the Particle Data Group (PDG), the lattice QCD calculations, and QMD/UrQMD simulations shall be presented. It intends not only to characterize the meson spectra in thermal and dense medium but to describe their vacuum phenomenology in finite magnetic fields. We conclude that the results obtained are remarkably precise, especially for some light mesons, if these are extrapolated to vanishing temperatures.

The arrangement of the present paper is as follows. We briefly describe the chiral LSM Lagrangian with three quark flavors in section \ref{lagrangian}. Then, we introduce a short reminder to the PLSM model in the mean-field approximation in section \ref{PolyakovLSM}. In vanishing and finite magnetic field, the dependence of chiral nonstrange  and strange quark condensates and the corresponding deconfinement order-parameters characterizing the quark-hadron phase transition on the temperature and the  chemical potential calculated form the SU($3$) PLSM model shall be presented in section \ref{Order:parameters}. The characterization of the magnetic catalysis shall be discussed, as well. The chiral phase-structure of the various meson states in finite magnetic field shall be outlined in section \ref{mesons}. The temperature dependence of the chiral phase-structure of sixteen meson states in finite magnetic field shall be analyzed in section \ref{sec:Tdepnd}. The chemical potential dependence shall be introduced in section \ref{sec:mudepnd}. Section \ref{sec:lMatsubaraF} is devoted to the temperature dependences of these meson states normalized to the lowest Matsubara frequencies. The conclusions are listed out in section \ref{conclusion}.

% -------------------------------------------------------------------------------
%                                       Model
%--------------------------------------------------------------------------------
\section{A short reminder to SU(3) Polyakov linear-sigma model \label{model}}

To study the QCD thermodynamics and characterize the various nonet mesons at finite temperatures and chemical potentials, the linear-sigma model shall be utilized as an effective QCD-like model. In this section, we shall briefly describe the chiral phase-structure of LSM Lagrangian with three quark flavors $N_f=3$. This can be established as fermionic and mesonic potentials. Concretely, the structure of this given Lagrangian shall provide us with tools to investigate the chiral phase-structure of nonet meson states. As we have done in a serious of previous publications, we treat SU($3$) PLSM by means of the mean-field approximation, despite its well-know constrains. This approach enables us to characterize respective influences of finite magnetic field on the quark-hadron phase transition in both thermal and dense medium. With dense medium we mean, even if it might not be exactly equivalent with, finite chemical potential. It intends to examine these for the chiral phase-structure of the sixteen meson states. Doing this we are able to compare their mass spectra with the experimental results, with the first-principle lattice QCD calculations and with the corresponding estimations from other QCD-like models.

\subsection{Lagrangian of linear-sigma model \label{lagrangian}}

As introduced in previous sections, LSM seems to give a remarkably excellent description for various meson states. With increasing the degrees of freedom, the quark flavors $N_f$, the number of meson states that can be generated, increases, as well. The full LSM Lagrangian in $U(N_f)$ with the color degrees of freedom $N_c$ is defined as $\mathcal{L}_{chiral}=\mathcal{L}_f+\mathcal{L}_m$, i.e. fermionic and mesonic sectors. In other words 
\begin{itemize}
\item the first term represents quarks ($q$) and antiquarks fields ($\bar{q}$) \cite{Tawfik:2014gga},
\begin{equation}
\label{eq:quarkL}
\mathcal{L}_{f} =\bar{q} \left[i \dslash - g\; T_a\,\left(\sigma_a + i\,  \gamma_5\, \pi_a + \gamma _{\zeta} V_a ^{\zeta}+\gamma _{\zeta}\gamma _{5} A_a ^{\zeta} \right)\,\right] q, 
\end{equation}
with $g$ is the Yukawa coupling constant and $\zeta$ is an additional Lorentz index \cite{Koch1997} and
\item the second term gives the mesonic Lagrangian, which is consisting of various contributions of the nonet states, the interactions and the possible anomalies
\bea
\mathcal{L}_{m}&=&\mathcal{L}_{SP}+ \mathcal{L}_{VA}+ \mathcal {L}_{Int}+ \mathcal{L}_{U(1)_A},  
\eea
where $\mathcal{L}_{SP}$ stand for scalars ($J^{pc}=0^{++}$) and pseudoscalars ($J^{pc}=0^{-+}$), while $\mathcal{L}_{AV}$ represent vectors ($J^{pc}=1^{--}$) and axial-vectors ($J^{pc}=1^{++}$) mesons. $\mathcal{L}_{Int}$ represent interactions that take place in the system of interest. The last term is called an anomaly term $U(1)_A$. For greater details about the mesonic Lagrangian, interested readers are kindly advised to consult Refs. \cite{Tawfik:2014gga,Gell Mann:1960,Gasiorowicz;1969, Rudaz:1994, Parganlija:2008, Pisarski:1995A,Pisarski:1995B,Pisarski:1995C, Wolf:anomaly}.  Further details about $U(N_f)_r \times U(N_f)_{\ell}$ can be taken from \cite{Gell Mann:1960,Gasiorowicz;1969,Rudaz:1994,Parganlija:2008,Pisarski:1995A,Pisarski:1995B,Pisarski:1995C,Wolf:anomaly}
\bea
\mathcal{L}_{SP}&=&\mathrm{Tr}\left[(D^{\mu}\Phi)^{\dag}\,(D^{\mu}\Phi)-m^2
\Phi^{\dag} \Phi\right]-\lambda_1 [\mathrm{Tr}(\Phi^{\dag} \Phi)]^2
-\lambda_2 \mathrm{Tr}(\Phi^{\dag}
\Phi)^2 + \mathrm{Tr}[H(\Phi+\Phi^{\dag})], \label{eq:scalar_nonets} \\ 
  \mathcal{L}_{AV}&=&-\frac{1}{4}\mathop{\mathrm{Tr}}(L_{\mu\nu}^{2}+R_{\mu\nu}^{2}
)+\mathop{\mathrm{Tr}}\left[  \left( \frac{m_{1}^{2}}{2}+\Delta\right)  (L_{\mu}^{2}+R_{\mu}^{2})\right] \nn \\
&+&i \frac{g_{2}}{2} (\mathop{\mathrm{Tr}}\{L_{\mu\nu}[L^{\mu},L^{\nu}]\}+\mathop{\mathrm{Tr}}\{R_{\mu\nu}[R^{\mu},R^{\nu}]\}){\nonumber}\\
&+& g_{3}[\mathop{\mathrm{Tr}}(L_{\mu}L_{\nu}L^{\mu}L^{\nu}
)+\mathop{\mathrm{Tr}}(R_{\mu}R_{\nu}R^{\mu}R^{\nu})]+g_{4}
[\mathop{\mathrm{Tr}}\left(  L_{\mu}L^{\mu}L_{\nu}L^{\nu}\right)
+\mathop{\mathrm{Tr}}\left(  R_{\mu}R^{\mu}R_{\nu}R^{\nu}\right)
]{\nonumber}\\
&+&g_{5}\mathop{\mathrm{Tr}}\left(  L_{\mu}L^{\mu}\right)
\,\mathop{\mathrm{Tr}}\left(  R_{\nu}R^{\nu}\right)  +g_{6}
[\mathop{\mathrm{Tr}}(L_{\mu}L^{\mu})\,\mathop{\mathrm{Tr}}(L_{\nu}L^{\nu
})+\mathop{\mathrm{Tr}}(R_{\mu}R^{\mu})\,\mathop{\mathrm{Tr}}(R_{\nu}R^{\nu
})],\label{eq:vector_nonets}
\\
  \mathcal{L}_{Int}&=&\frac{h_{1}}{2}\mathop{\mathrm{Tr}}(\Phi^{\dagger}\Phi
)\mathop{\mathrm{Tr}}(L_{\mu}^{2}+R_{\mu}^{2})+h_{2}%
\mathop{\mathrm{Tr}}[\vert L_{\mu}\Phi \vert ^{2}+\vert \Phi R_{\mu} \vert ^{2}]+2h_{3}%
\mathop{\mathrm{Tr}}(L_{\mu}\Phi R^{\mu}\Phi^{\dagger}),\label{eq:INT}
\\ 
\mathcal{L}_{U(1)_A}&=&c[\mathrm{Det}(\Phi)+\mathrm{Det}(\Phi^{\dag})]+c_0 [\mathrm{Det}(\Phi)-\mathrm{Det}(\Phi^{\dag})]^2 +c_1 [\mathrm{Det}(\Phi)+\mathrm{Det}(\Phi^{\dag})]\,\mathrm{Tr} [\Phi \Phi^{\dag}]. \hspace*{10mm}
 \label{eq:Lagrangian}
\eea
\end{itemize}

%Table \ref{tab:1a} summarizes the various parameters of Eqs. (\ref{eq:scalar_nonets})-(\ref{eq:Lagrangian}).

%%%%%%%%%%%%%%%%%%%%%%%%Table
%\begin{table}[htb]
%\begin{center}
%\begin{tabular}{|c | c | c | c | c | c | c | c|}
%\hline 
%PLSM parameters & $c\,$ [MeV] & $h_x\,$ [MeV$^3$] & $h_y\,$ [MeV$^3$] & $m^2 \,$ [MeV$^2$] & $\lambda _1$ & $\lambda _2$ & $g$\\ 
%\hline   \hline
%Finite anomaly term \cite{Schaefer:2009} & $4807.84$ & $(120.73)^3$ & $(336.41)^3$ & -$(306.26)^2$ & $13.49$& $46.48$&$6.5$\\ 
%Vanishing anomaly term \cite{Schaefer:2009} & $0$ & $(120.73)^3$ & $(336.41)^3$& $-(503.55)^2$ & -$4.55$ &$82.47$ &$6.5$\\ 
%\hline  
%(Axial)vector \cite{Rischke:2012} & $h_1$ & $h_2$ & $h_3$ & $m_1^2$ [MeV$^2$] & $\delta _x$ [MeV$^2$] &  $\delta_y$ [MeV$^2$] & $g_1$\\ \hline 
% & $0$ & $9.87$ & $4.8667 $ & $(0.4135)^2$ & $0$ & $(0.1511)^2$&$6.5$ \\ 
%\hline 
%\end{tabular}
%\caption{A summary of the various PLSM parameters that should be fixed and enter our calculations. \label{tab:1a}}
%\end{center}
%\end{table} 
%%%%%%%%%%%%%%%%%%%%%%%%%%%%%%%%%%%%%%%%%%%%%
 
The {\it complex} matrices for scalars $\sigma_{a}$, i.e. $J^{PC}=0^{++}$, pseudoscalars  $\pi _{a}$, i.e. $J^{PC}=0^{-+}$, vectors $V_{a}^{\mu}$, i.e. $J^{PC}=1^{--}$ and axial-vectors $A_{a}^{\mu}$, i.e. $J^{PC}=1^{++}$ meson states can be constructed as
\bea
\label{fieldmatrix}
\Phi 	 = \sum_{a=0}^{N_{f}^{2} -1} T_{a}(\sigma _{a}+ i \pi _{a}), \qquad  \qquad
L^{\mu}  = \sum_{a=0} ^{N_{f}^{2} -1} \, T_{a}\, (V_{a}^{\mu}+A_{a}^{\mu}), \qquad  \qquad
R^{\mu}  = \sum_{a=0}^{N_{f}^{2} -1}\, T_{a}\, (V_{a}^{\mu}-A_{a}^{\mu}).
\eea
The various generators are defined according to $N_f$. It is obvious that the covariant derivative, $D^\mu \Phi = \partial^\mu \Phi-i\,g_1 (L^\mu \Phi - \Phi R^\mu)$, is to be associated with the degrees of freedom for  (pseudo-)scalar  and (axial-)vector and couples them through  $g_1$, the coupling constant. For instance, for $N_f=3$, the unitary matrices for $SU(3)_r \times SU(3)_{\ell}$ can be taken from Ref. \cite{Tawfik:2014gga}, while $T_{a}$ the generators of $U(3)$ can be expressed as $T_{a}=\hat{\lambda}_{a}/2$, with $a=0\dots 8$ and $\hat{\lambda}_{a}$ are Gell-Mann matrices.

Different PLSM parameters have been fixed in previous studies \cite{Schaefer:2009,Rischke:2012}. It should be noticed that these two models are inconsistent and so does any parameterization based on them. The problem was solved in Ref.  \cite{Rischke:2004}, where the authors introduced $C_1$ combining various parameters $C_1=m_0^2+\lambda_1(\sigma_l^2+\sigma_s^2)$. While $C_1$ can only be estimated through parametrization, its individual parameters do not. While both $m_0^2$ and  $\lambda_1$ were taken from Ref. \cite{Schaefer:2009}, the remaining ones have been determined (see Tab. III in Ref. \cite{Rischke:2012}).

\subsection{Polyakov linear-sigma model in mean-field approximation \label{PolyakovLSM}}
   
The LSM Lagrangian with $N_f=3$, which are coupled to $N_c=3$ can be given as $\mathcal{L}=\mathcal{L}_{chiral}-\mathbf{\mathcal{U}}(\phi, \phi^*, T)$. The potential $\mathbf{\mathcal{U}}(\phi, \phi^{*},T)$ should be adjusted from recent lattice QCD simulations. Accordingly, various LSM-parameters can be determined. It is should be noticed that this Lagrangian has $Z(3)$ center symmetry \cite{Ratti:2005jh,Roessner:2007,Schaefer:2007d,Fukushima:2008wg}. Through the thermal expectation values of the color traced Wilson-loop in the temporal direction, the dynamics of color charges and gluons can be taken into consideration
\begin{eqnarray}
\phi = \langle\mathrm{Tr}_c\, \mathcal{P}\rangle/N_c, \qquad && \qquad
\phi^* = \langle\mathrm{Tr}_c\,  \mathcal{P}^{\dag}\rangle/N_c. \label{phis}
\end{eqnarray}
The Polyakov-loop potential $\mathbf{\mathcal{U}}(\phi, \phi^*, T)$ can be introduced in the pure gauge limit as a temperature- and a density-dependent quantity. There are various proposals introduced, so far. In

In previous works, we have implemented polynomial potential \cite{Tawfik:2014uka,Tawfik:2014gga,Tawfik:2014hwa,Tawfik:2014bna}. In the present work, we introduce calculations based on an alternatively-{\it improved} extension to $\phi$  and $\phi^{*}$ \cite{Roessner:2007,Sasaki:2013ssdw,Fukushima:2008wg}; the logarithmic potential, 
\bea
\frac{\mathbf{\mathcal{U}}_{\mathrm{Log}}(\phi, \phi^*, T)}{T^4} = \frac{-a(T)}{2} \; \phi^* \phi + b(T)\; \ln{\left[1- 6\, \phi^* \phi + 4 \,( \phi^{*3} + \phi^3) - 3 \,( \phi^* \phi)^2 \right]}. \label{LogULoop}
\eea 
where with $T_0$ being critical temperature for deconfinement phase-transition in the pure-gauge sector, 
\bea
a(T) = a_0 + a_1 \left(T0/T\right) + a_2 \left(T_0/T\right)^2  \qquad  \mathrm{and}  \qquad b(T) =
 b_3  \left(T_0/T\right)^3,
\eea
The remaining parameters $a_0$, $a_1$, $a_2$, and $b_3$ can be determined through confrontation with lattice QCD simulations. These are listed in Tab. II in Ref.  \cite{Tawfik:2016edq}.

In thermal equilibrium, the mean-field approximation of the Polyakov linear-sigma model can be implemented in the grand-canonical partition function ($\mathcal{Z}$) at finite temperature ($T$) and finite chemical potential ($\mu_f$). Here, the subscript $f$ refers to the quark flavors. At finite volume ($V$), the free energy reads $\mathcal{F}=-T \cdot \log [\,\mathcal{Z}]/V$. For instance, for SU($3$) PLSM, 
\begin{equation}
\mathcal{F} =  U(\sigma_l, \sigma_s) + \mathbf{\mathcal{U}}(\phi, \phi^*, T) + \Omega_{\bar{q}q} (T, \mu_f, B). \label{potential}
\end{equation}
These three terms (potentials) can be elaborated as follows.
\begin{itemize}
\item The purely mesonic potential, the first term, stands for strange ($\sigma_s$) and nonstrange ($\sigma_l$)  condensates, i.e. 
 \begin{eqnarray}
U(\sigma_l, \sigma_s) &=& - h_l \sigma_l - h_s \sigma_s + \frac{m^2}{2}\, (\sigma^2_l+\sigma^2_s) - \frac{c}{2\sqrt{2}} \sigma^2_l \sigma_s  \nonumber \\ 
&+&  \frac{\lambda_1}{2} \, \sigma^2_l \sigma^2_s +\frac{(2 \lambda_1 +\lambda_2)}{8} \sigma^4_l  + \frac{(\lambda_1+\lambda_2)}{4}\sigma^4_s . \hspace*{8mm} \label{Upotio}
\label{pure:meson}
\end{eqnarray}

\item The Polyakov-loop potential, the second term, was already detailed in Eq. (\ref{LogULoop}).

\item The quarks and antiquark potential, the third term, is obviously a subject of remarkable modifications through finite external magnetic field, scope of the present work,  
\begin{eqnarray} 
\Omega_{  \bar{q}q}(T,\;\mu_f,\;B)&=& - 2 \sum_{f} \frac{|q_f| B \, T}{(2 \pi)^2} \,  \sum_{\nu = 0}^{\infty}  (2-\delta _{0 \nu })    \int_0^{\infty} dp_z \nonumber \\ && \hspace*{5mm} 
\left\{ \ln \left[1+3\left(\phi+\phi^* e^{-\frac{E_{B, f} -\mu_f}{T}}\right)\; e^{-\frac{E_{B, f} -\mu_f}{T}} +e^{-3 \frac{E_{B, f} -\mu_f}{T}}\right] \right. \nonumber \\ 
&& \hspace*{3.7mm} \left.+\ln \left[ 1+3\left(\phi^*+\phi e^{-\frac{E_{B, f} +\mu_f}{T}}\right)\; e^{-\frac{E_{B, f} +\mu_f}{T}}+e^{-3 \frac{E_{B, f} +\mu_f}{T}}\right] \right\}. \label{PloykovPLSM}
\end{eqnarray}
$E_{B,f}$ is the dispersion relation of each of the quark flavors in finite external magnetic field, $B\neq 0$
\bea
E_{B, f}&=&\sqrt{p_{z}^{2}+m_{f}^{2}+|q_{f}|(2n+1-\Sigma) B}, \label{eq:moddisp}
\eea 
where $\sigma$ is represent the spin quantum number, $\sigma=\pm S/2$ and $m_{f}$ are the masses of quark flavors, which are directly coupled to the corresponding sigma-fields,
\bea
m_l = g\, \frac{\bar{\sigma_l}}{2}, \qquad & & \qquad
m_s = g\, \frac{\bar{\sigma_s}}{\sqrt{2}}.  \label{qmassSigma}
\eea
For the sake of completeness, we recall that the light and strange condensates are to be determined from the partially conserved axial-vector current (PCAC) relation \cite{Schaefer:2009},
\bea
\bar{\sigma_l}= f_{\pi},  \qquad & & \qquad \bar{\sigma_s} \frac{1}{\sqrt{2}}(2f_K-f_\pi).
\eea
with $f_{\pi}$ and $f_K$ are pion and kaon decay constants, respectively. The quantity $2-\delta_{0 \nu}$ is related to the degenerate Landau Levels $\nu$. We like to highlight that, the quantity $2n+1-\sigma$ can be replaced by a summation over the Landau Levels ($\nu$). In nonzero magnetic field  ($eB\neq 0$) but finite $T$ and $\mu_f$,  both Landau quantization and magnetic catalysis, where the magnetic field is assumed to be oriented along $z$-direction, have been implemented. The quantization number ($n$) is known as the Landau quantum number. For details about Landau Levels and how they are occupied, interested readers are kindly advised to consult Ref. \cite{Tawfik:2016ihn}.
\end{itemize}

Regarding to the fermionic vacuum term let us now review the status of its role in various QCD-like models. In Ref. \cite{Redlich:2010d}, it was concluded that the inclusion of the fermionic vacuum term commonly known as no-sea approximation seems to cause a second-order phase transition in the chiral limit. Accordingly, it was widely believed that its inclusion is accompanied with first- or second-order phase transition depending on baryon chemical potential but also on the choice of the coupling constants \cite{Redlich:2010d}. For a further investigation of its role, various thermodynamic observables have been evaluated in two different effective models, NJL and quark-meson (QM) models or LSM. To this end, the adiabatic trajectories in the QM model are found exhibiting a {\it kink} at the chiral phase transition, while in the NJL model, this was a {\it smooth} transition everywhere \cite{blind}. In light of this, it was suggested that the fermionic vacuum term can be dropped out in the QM model, similar to the approach utilized in the present work, while in the NJL model, this term must be included. This effect underlies a first-order phase transition in the LSM model in chiral limit, especially when the fermionic vacuum fluctuations can be neglected \cite{Redlich:2010as}. The direct method of removing the ultraviolet divergences is the one in which the fermionic vacuum term is included \cite{Menezes:2009a,Menezes:2009a},
\bea
\Omega^{\mbox{vac}}_{q\bar{q}} &=&  2 N_c N_f \sum_f \int_\Lambda \frac{d^3p}{(2\pi)^3} E_f = \frac{-N_c N_f}{8\pi^2} \sum_f \left( m_f^4 \ln {\left[\frac{\Lambda + \epsilon_\Lambda}{m_f} \right]} - \epsilon_\Lambda \left[\Lambda^2 + \epsilon_\Lambda^2\right]\right),
\eea 
where $\epsilon_\Lambda=(\Lambda^2 + m_f^2)^{1/2}$ and $m_f$ is the flavor mass. Eq. (\ref{PloykovPLSM}) gives the fermionic contributions to the medium in finite magnetic field, Therefore, many authors think that the fermionic vacuum term has a negligible effect on the PLSM results. So we assume.

Coming back to our approach of PLSM and its mean field approximation, we recall that the mean values of PLSM order parameters such as the chiral condensates $\sigma_l$ and $\sigma_s$ and deconfinement phase transitions $\phi$ and $\phi^*$, are  evaluated in the way that one minimizes the free energy $\mathcal{F}$ at finite volume ($V$) with respect the corresponding filed, where  $\sigma_l=\bar{\sigma_l}$, $\sigma_s=\bar{\sigma_s}$, $\phi=\bar{\phi}$ and $\phi^*=\bar{\phi^*}$,
\begin{eqnarray}
\left.\frac{\partial \mathcal{F} }{\partial \sigma_l} = \frac{\partial
\mathcal{F}}{\partial \sigma_s}= \frac{\partial \mathcal{F} }{\partial
\phi}= \frac{\partial \mathcal{F} }{\partial \phi^*}\right|_{min} &=& 0. \label{cond1}
\end{eqnarray}
Their dependence on $T$, $\mu$ and $e B$ can then be determined. At vanishing chemical potential ($\mu=0$), it is obvious that the order parameters of the deconfinement phase transitions, $\phi$ and $\phi^*$, are identical, but at $\mu \ne 0$, they are distinguishable. The PLSM free energy at finite $V$, Eq. (\ref{potential}), becomes complex. Therefore, the analysis of PLSM order parameters should be given by minimizing the real part of free energy $Re(\mathcal{F})$ at the  saddle point.

%--------------------------------------------------------------------------------
%                                        Mganetic properties
%--------------------------------------------------------------------------------
\section{Results \label{resulat}}

We study the dependence of the quark-hadron phase transition on temperature and baryon chemical potential in presence of a finite magnetic field. To this end, we first estimate the respective influences of the magnetic field on the various chiral condensates and the deconfinement order-parameters, Then, in the second part of the present script, we estimate the chiral phase-structure of sixteen meson states in thermal and dense  medium. With dense medium, we mean medium with finite chemical potential. Last but not least, we shall present temperature  and chemical potential dependences of the sixteen meson states normalized  to the lowest Matsubara frequency, as well.

\subsection{Order parameters and magnetic catalysis} 
\label{Order:parameters}

The chiral condensates ($\sigma_l$ and $\sigma_s$) and the deconfinement order-parameters ($\phi$ and $\phi^*$) shall be analyzed at a wide range of temperatures, baryon chemical potentials, and magnetic field strengths.  The various PLSM parameters are estimated at the vacuum mass sigma meson $\sigma=800~$MeV, where the vacuum nonstrange and strange chiral condensates are $\sigma_{l_o}=92.5~$ MeV and $\sigma_{s_o}=94.2~$MeV, respectively.

%----------------------- Fig. 1 order parameter f(T) eB= 1, 10 ------------------
\begin{figure}[htb]
\centering{
\includegraphics[width=5.cm,angle=-90]{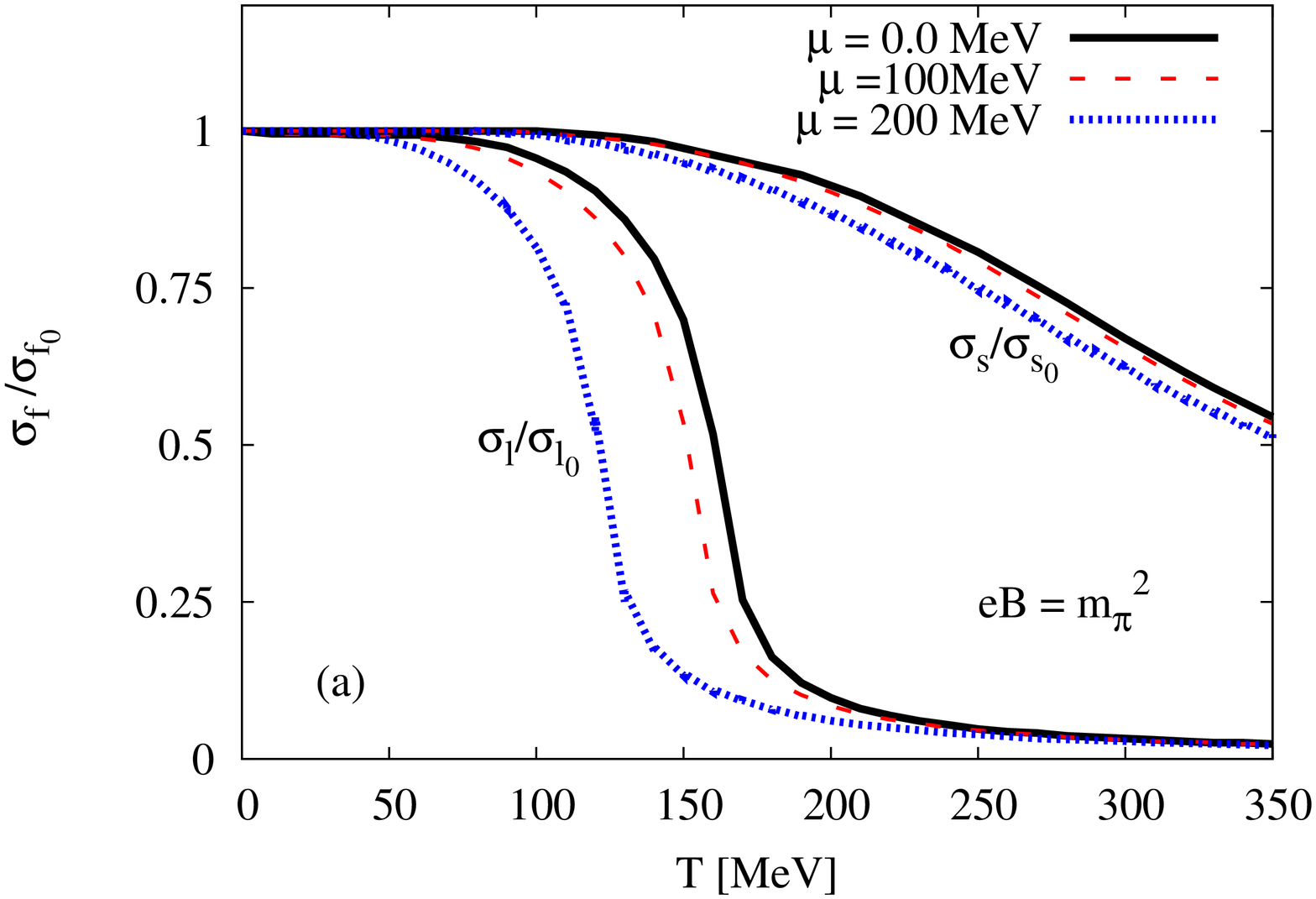}
\includegraphics[width=5.cm,angle=-90]{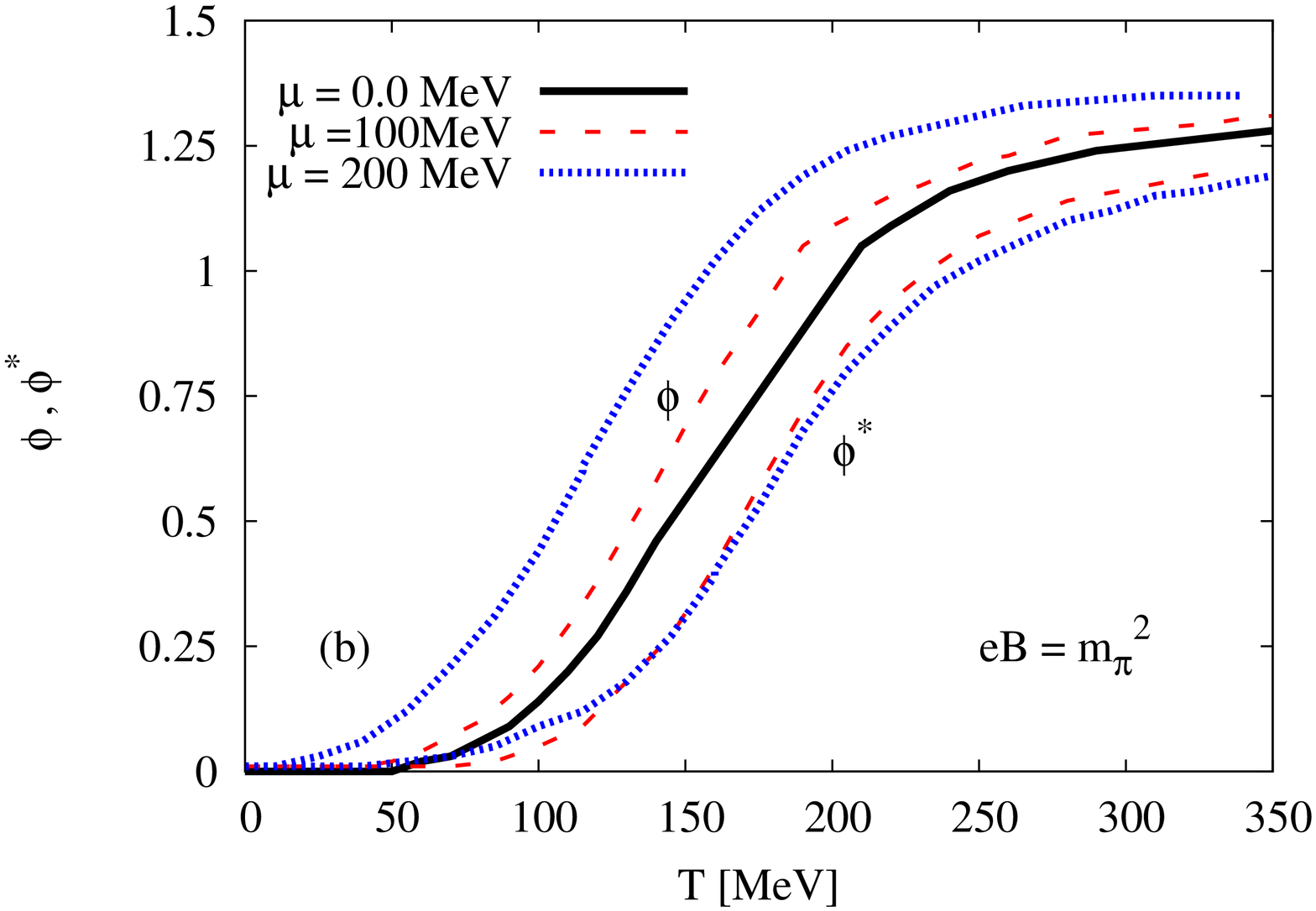}\\
\includegraphics[width=5.cm,angle=-90]{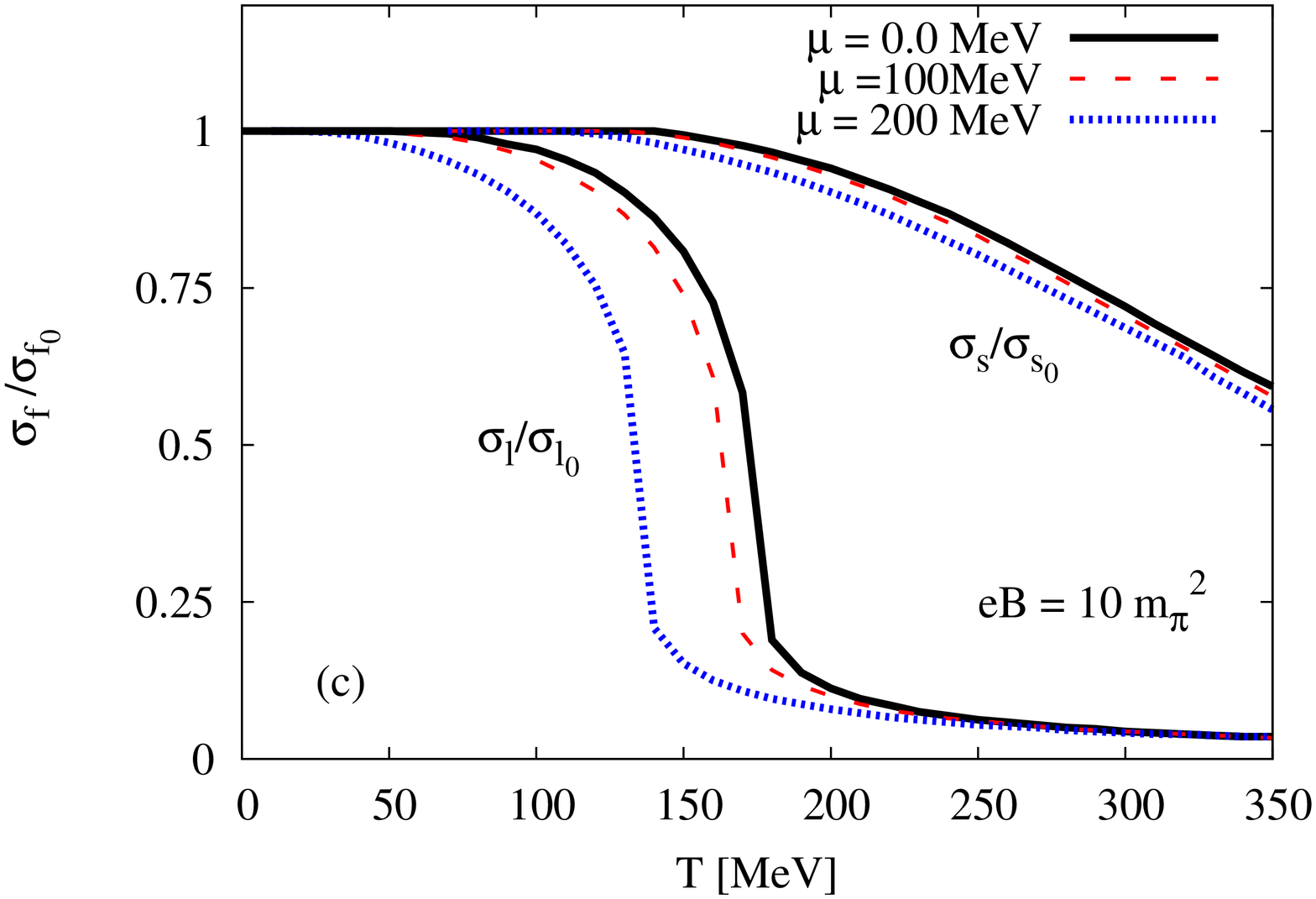}
\includegraphics[width=5.cm,angle=-90]{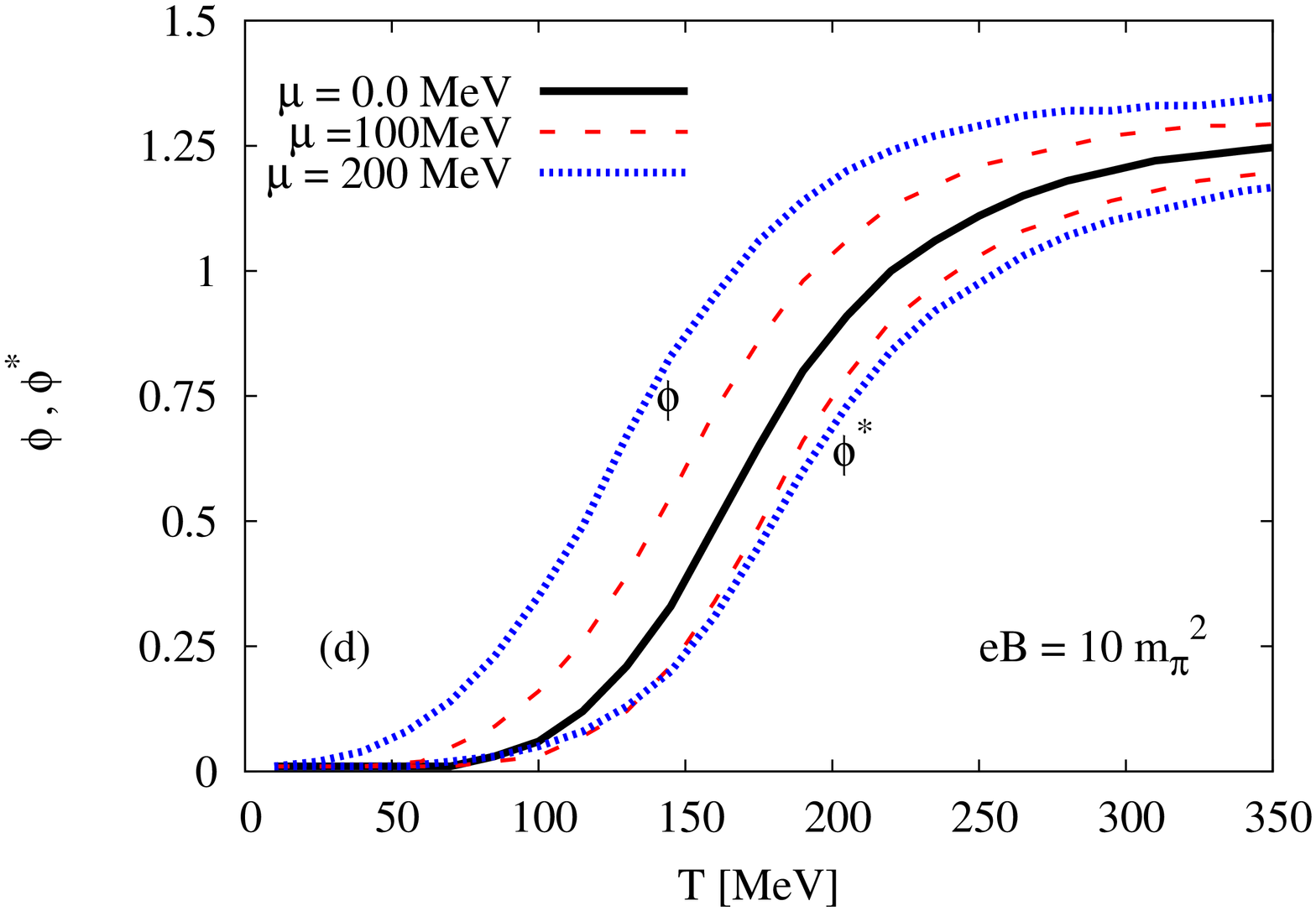}
\caption{(Color online)  Left-hand panels (a) and (c) show normalized chiral-condensate with respect to the vacuum value as functions of temperature. Right-hand panels (b) and (d) give the expectation values of the Polyakov-loop fields ($\phi$  and $\phi^*$) as functions of temperature. The upper panels present results at magnetic fields $eB =m^2_{\pi}~$, while bottom panels at $eB=10 \,m^2_{\pi}~$. The calculations are performed at different baryon chemical potentials; $\mu=0$ (solid curves), $100$ (dashed curves), $200~$MeV (dotted curves).  \label{fig:sbtrc1}}}
\end{figure}

In Fig. \ref{fig:sbtrc1}, the temperature dependences of the different order parameters are calculated at different magnetic fields;  $eB=m^2_{\pi}~$ (upper panels) and  $eB=10\, m^2_{\pi}~$(bottom panels) and at different baryon chemical potentials; $\mu=0$ (solid curves), $100$ (dashed curves), $200~$MeV (dotted curves).  In left-hand panels (a)  and (c), the normalized chiral condensates ($\sigma_{l}/\sigma_{l_o}$ and $\sigma_{s}/\sigma_{s_o}$) are depicted. The upper panels introduce results at magnetic fields $eB =m^2_{\pi}~$, while the bottom panels at $eB=10 \,m^2_{\pi}~$. The dependences on the baryon chemical potentials are also depicted. We illustrate results at $\mu=0$ (solid curves), $100$ (dashed curves), and $200~$MeV (dotted curves). We notice that the chiral condensates are slightly shifted to lower critical temperatures with increasing the baryon chemical potential. In other words, the critical chiral temperature ($T_{\chi}$) decreases with the increase in the magnetic field ($eB$) and with the increasing in $\mu$, as well.  The procedure utilized in determining $T_{\chi}$ shall be elaborated later. Such as observation is known as inverse magnetic catalysis. We bear in mind that this kind of inverse magnetic catalysis is related to the baryon chemical potential. This might be seen as a novel discovery to be credited to the present work. Furthermore, we can conclude that the effect of strong magnetic field lays in the same line as that of the baryon chemical potentials, especially on affecting earliness of the phase transition, i.e. that the phase transition takes place at lower temperatures. In other words, similar to the inverse magnetic catalysis corresponding to the temperature, we also observed  an inverse magnetic catalysis related to magnetic field, as well.

Right-hand panels (b) and (d) draw the Polyakov-loop order parameters ($\phi$ and it's conjugation $\phi^*$)  as functions of temperature at same values of baryon chemical potentials and magnetic fields as in the left-hand panels. It is obvious that at $\mu=0~$MeV (solid curves), $\phi=\phi ^*$, i.e. both order parameters aren't distinguishable. But they become different at finite $\mu$; $\mu=100~$MeV (dashed curves) and $\mu= 200~$MeV (dotted curves). The effects of the magnetic field are obvious. We observe that the deconfinement critical temperature ($T_{\phi}$) is shifted to lower values as the magnetic field increase and as well as the baryon chemical potential increase. This is similar to the left-hand panels. On the other hand, we conclude that $\phi$ and $\phi^*$ have different dependences on the temperature. While $\phi$ increases as the magnetic field increase and as the baryon chemical potential increase, we find that $\phi^*$ decreases. Accordingly, the magnetic catalysis related to temperature is obviously either direct of inverse, respectively. Again, the determination of $T_{\phi}$ shall be discussed in a forthcoming section.

\begin{figure}[htb]
\centering{
\includegraphics[width=5.cm,angle=-90]{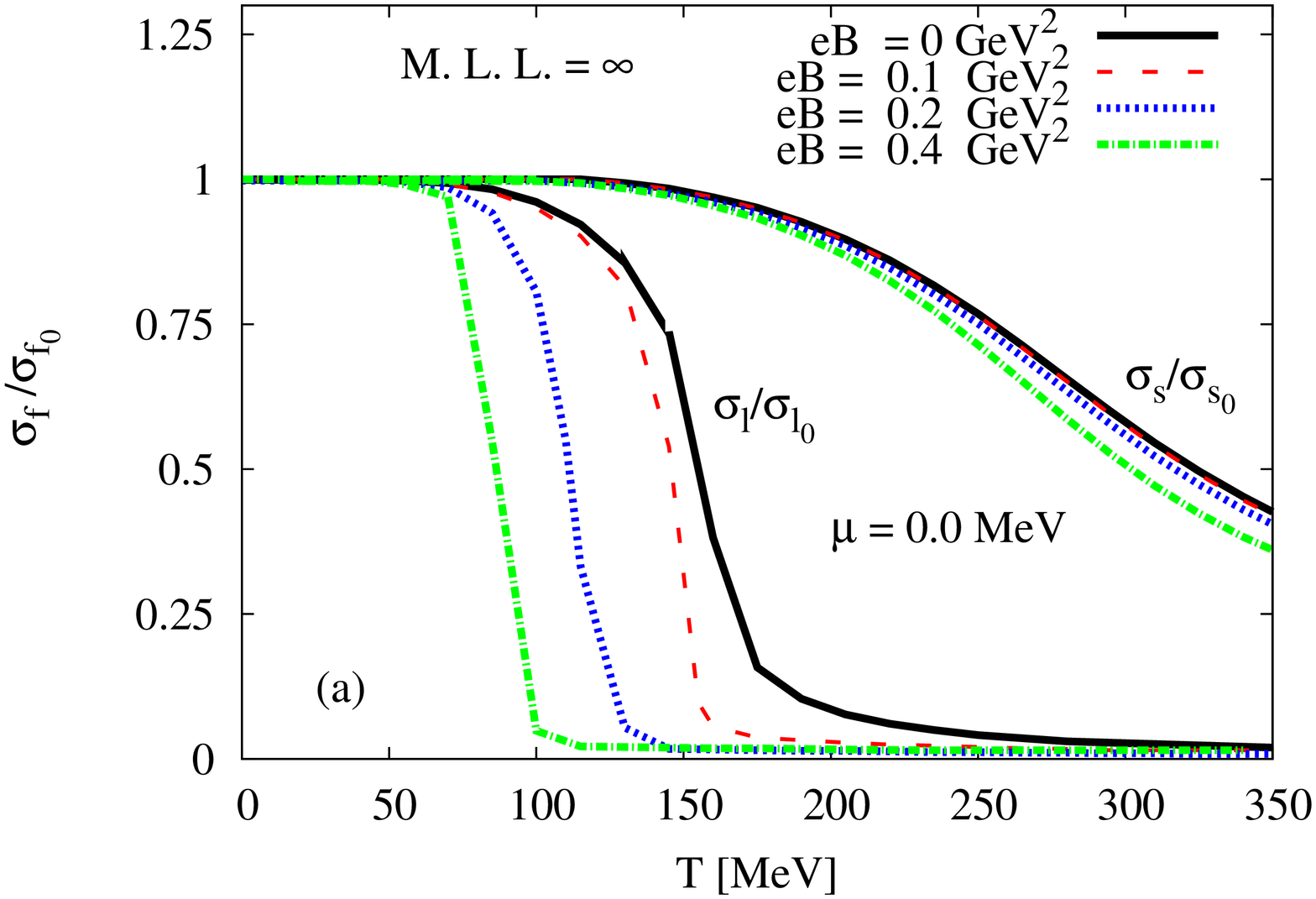}
\includegraphics[width=5.cm,angle=-90]{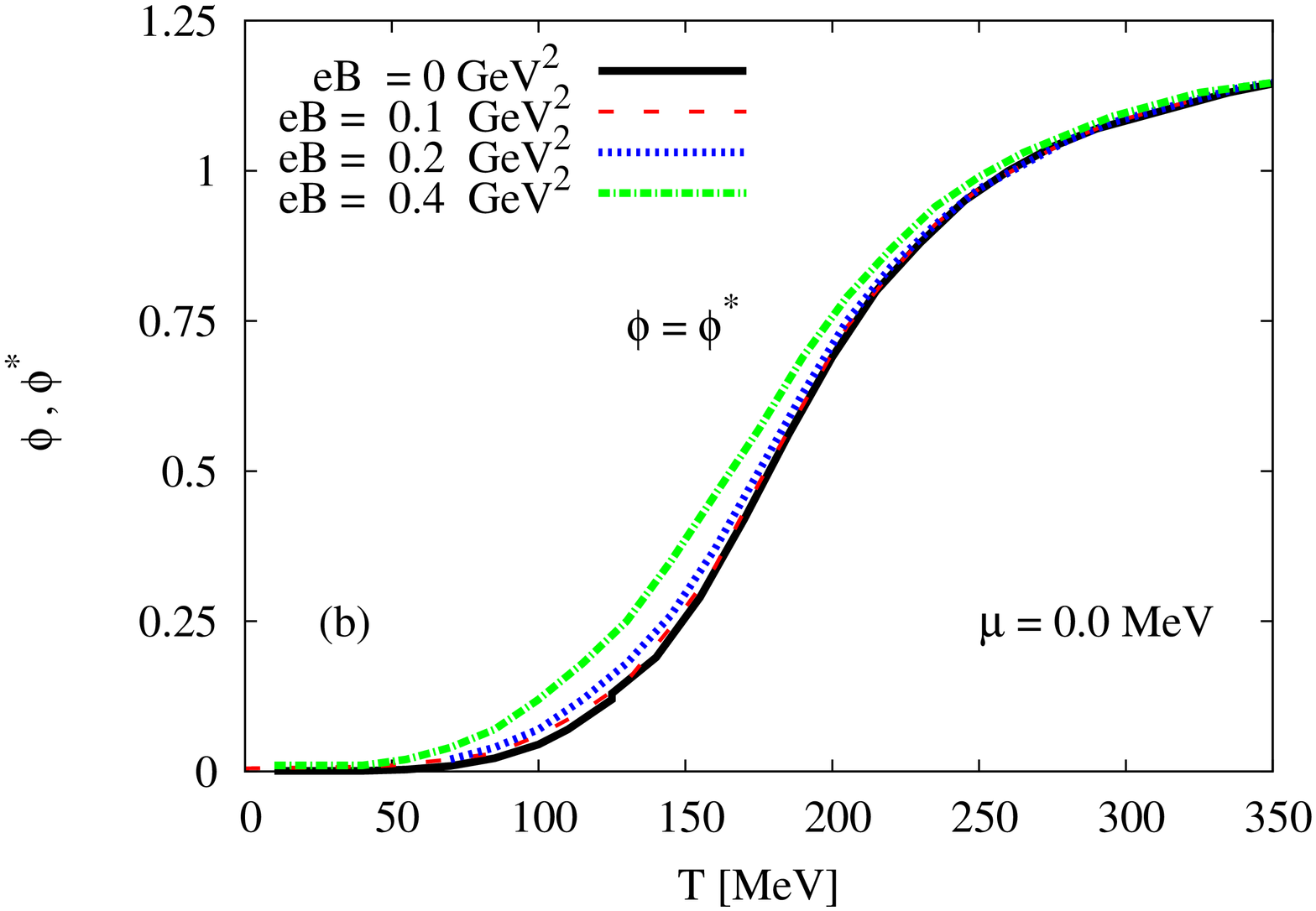}
\caption{(Color online)  Left-hand panel: the normalized chiral condensates are given as functions of temperature at vanishing baryon chemical potential $\mu=0$ but different magnetic fields; $eB=0$ (solid curves), $0.1$ (dashed curves), $0.2$ (dotted curves) and $0.4~$GeV$^2$ (dot-dashed curves). Right-hand panel: the same as in the left-hand panel but for the deconfinement order parameters $\phi$  and $\phi^*$. \label{fig:sbtrc2}
}}
\end{figure} 

Figure \ref{fig:sbtrc2} illustrates the temperature dependence of the normalized chiral quark-condensates (a) and deconfinement  order parameters (b) at different magnetic fields, $eB=0$ (solid curves), $0.1$ (dashed curves), $0.2$ (dotted curves) and $0.4~$GeV$^2$ (dot-dashed curves) and at vanishing chemical potential ($\mu=0$). Relative to the previous figure, this one presents a systematic study of the influences of the magnetic fields. In left-hand panel (a), we find that the critical chiral temperature deceases with increasing the magnetic field. This means that the phase transition, which is a crossover, becomes slightly sharper with increasing the magnetic field. In right-hand panel (b), the temperature dependence of the deconfinement order parameters is depicted at a vanishing baryon chemical potential, i.e. $\phi=\phi^*$, but different magnetic field strengths; $eB=0$ (solid curves), $0.1$ (dashed curves), $0.2$ (dotted curves) and $0.4~$GeV$^2$ (dot-dashed curves). It is obvious that the critical deconfinement temperature ($T_{\phi}$) very slightly decreases as the magnetic field increases.
 
In both figures, we assure that the maximum Landau levels are fully occupied with quark states. Great details about the Landau quantization as a crucial consequence of the magnetic field and they are accommodated with quark states can be taken from Refs. \cite{Tawfik:2016ihn,Tawfik:2017cdx}.

%-----------------------------------------Fig. 4 order parameter f(mu) T=50  and T = 100------------------
%
\begin{figure}[htb]
\centering{
\includegraphics[width=5.cm,angle=-90]{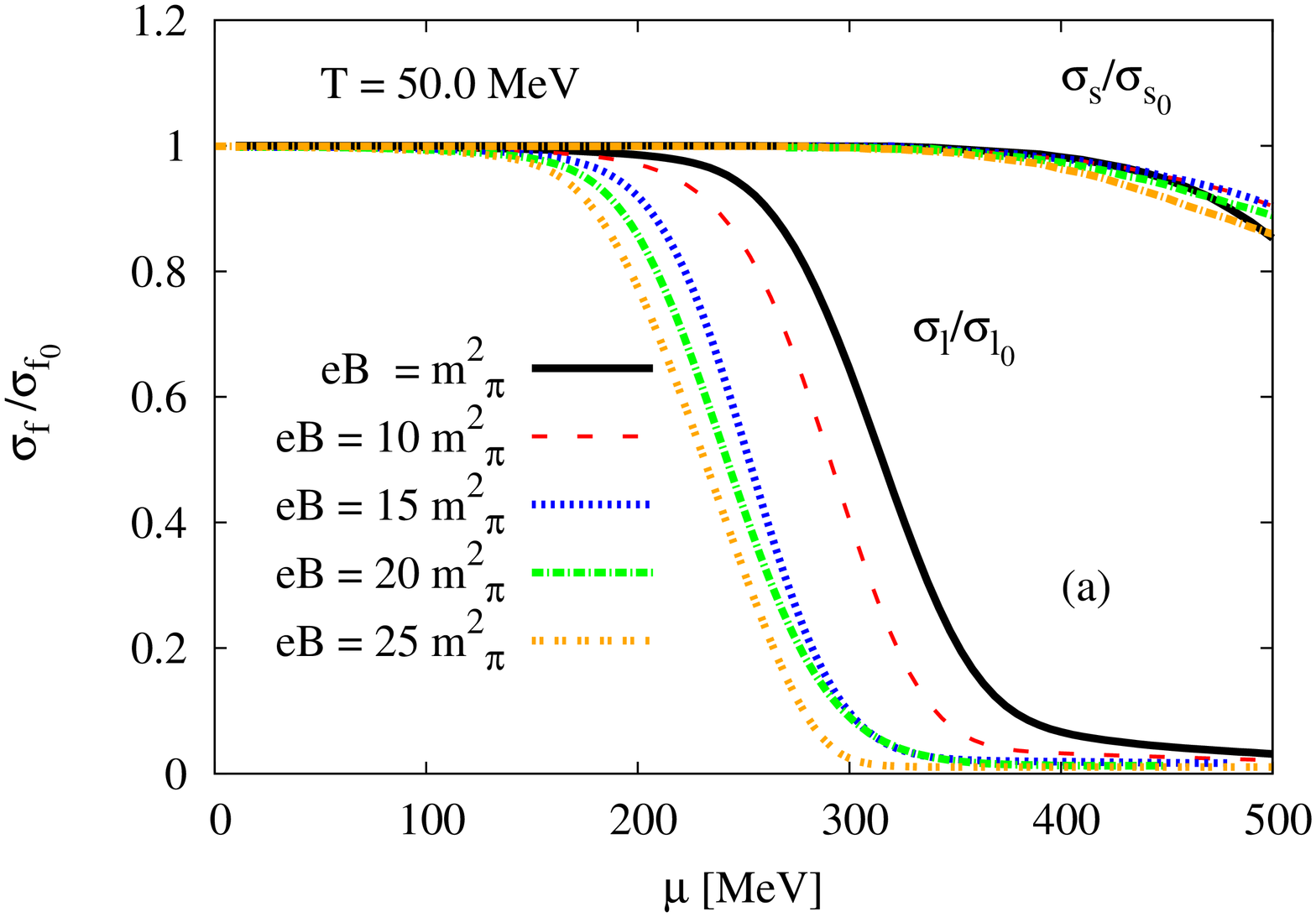}
\includegraphics[width=5.cm,angle=-90]{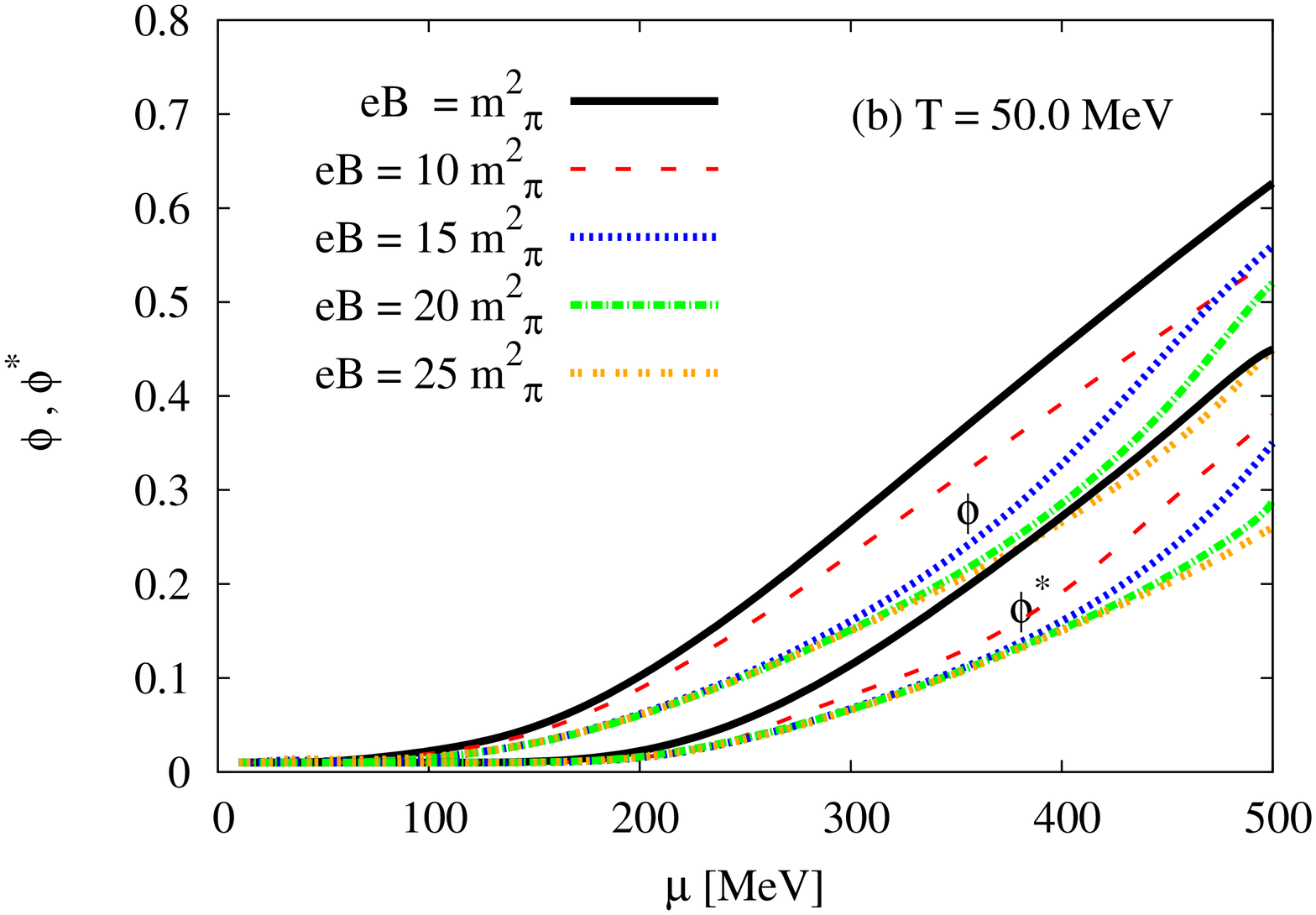}\\
\includegraphics[width=5.cm,angle=-90]{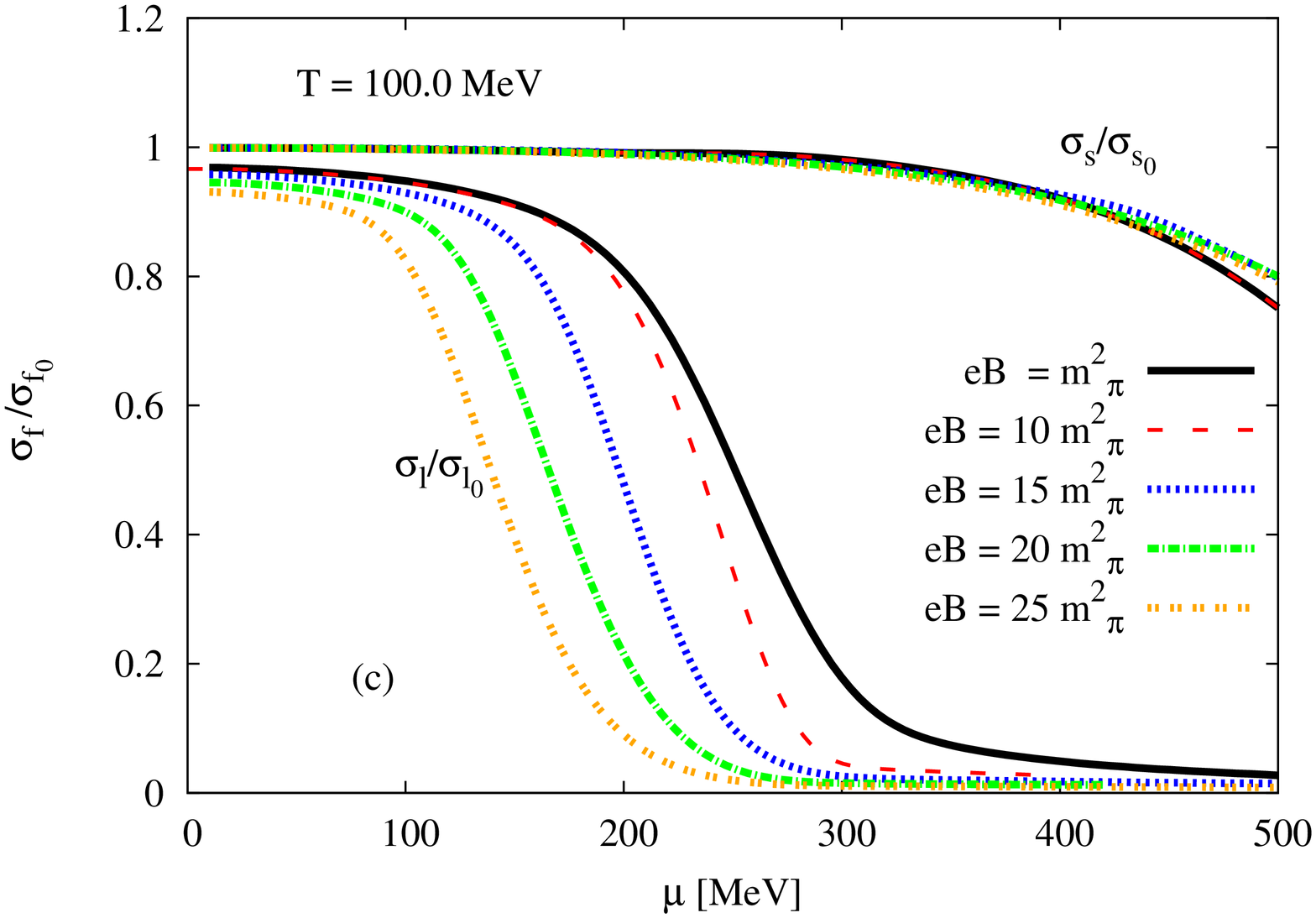}
\includegraphics[width=5.cm,angle=-90]{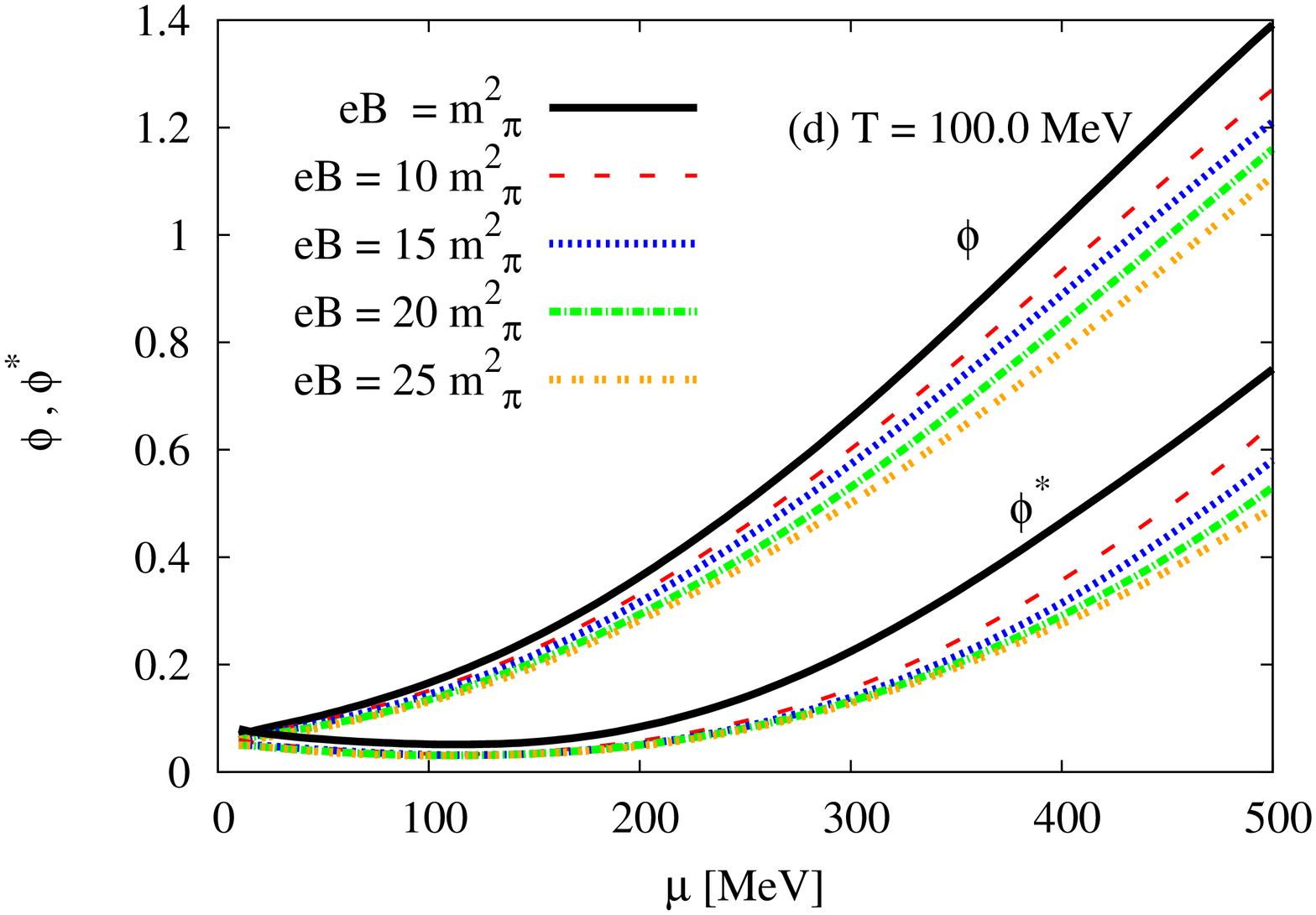}
\caption{(Color online) Left-hand panels (a) and (c) show the chiral condensates ($\sigma_l$ and $\sigma_s$) normalized  to the vacuum value as functions of the baryon chemical potential at different values of magnetic fields; $eB=1$ (solid curves) $10$ (dashed curves), $15$ (dotted curves), $20$ (dot dashed curves), and $25\,m^2_{\pi}~$ (double dotted curves). Right-hand panels (b) and (d) give the same as in the left-hand panels but for the expectation values of the Polyakov-loop potentials, i.e. deconfinement order parameters ($\phi$ and $\phi^*$). Top panels stand for results at $T=50\,$MeV, while bottom panels at $T=100\,$MeV. \label{fig:sbtrcMu}
}}
\end{figure}

Furthermore, we have calculated the chiral condensates ($\sigma_l$ and $\sigma_s$) and the deconfinement order-parameters ($\phi$  and $\phi^*$) in dependence on the baryon chemical potential ($\mu$) at different temperatures and magnetic field strengths. This is illustrated in Fig. \ref{fig:sbtrcMu}. At different temperatures; $T=50~$MeV (upper panels) and $T=100~$MeV (bottom panels), we complete the picture about the magnetic effects on the density (chemical potential) dependence of the four quantities; the chiral and deconfinement order parameters, respectively. In left-hand panels (a) and (c), the density dependence of the chiral condensates is given as functions of different magnetic field strengths; $eB=1$ (solid curves) $10$ (dashed curves), $15$ (dotted curves), $20$ (dot dashed curves), and $25\,m^2_{\pi}~$ (double dotted curves). We observe that increasing the temperatures causes a rapid decrease in the chiral condensates around the chiral phase transition similar to the ones observed in a previous study from PLSM without magnetic field \cite{Tawfik:2014gga} and in the previous two figures, as well. Such a decrease likely refers to a rapid crossover.

We observe that increasing the magnetic field sharpens the rapid decrease in the chiral phase-structure. This leads to a decrease in the corresponding {\it critical} chemical potential, which is be determined in a similar manner as the critical chiral temperature. Increasing the magnetic field tends to sharpen the phase transition, i.e. to accelerate the formation of a metastable phase characterizing the region of crossover. This phenomenon is known as a magnetic catalysis and actually an inverse one.

There is a gap difference observed between light and strange chiral condensates at high densities. This can be understood because of the inclusion of the anomaly term in Eq. (\ref{pure:meson}) which was discussion in section \ref{PolyakovLSM} and known as the $c$ term. The resulting fit parameters are accordingly modified \cite{Schaefer:2009,Tawfik:2014gga}. This was conjectured as an evident about the numerical estimation of the chiral condensates. The difference between $\sigma_l$ and $\sigma_s$ was also observed in the previous two figures.

The right-hand panels (b) and (d) give the deconfinement order parameter in dependence on baryon chemical potential at different temperatures; $T=50$ (top panel) and $T=100~$MeV (bottom panel) and fixed magnetic field strengths; $eB=1$ (solid curves) $10$ (dashed curves), $15$ (dotted curves), $20$ (dot dashed curves), and $25\,m^2_{\pi}$ (double dotted curves). The increase in the temperature increases the deconfinement order parameters to a larger baryon chemical potential. 

It is obvious that the slope of $\phi$ and $\phi^*$ with respect to $\mu$ can be approximately estimated. We observe that the magnetic field decreases their slopes, i.e. increases the corresponding critical chemical potentials, while increasing the temperature increases the slopes and thus decreases the corresponding critical temperatures. Both thermal and magnetic effects of the hadronic medium on the evolution of Polyakov-loop parameters seem to be very smooth.

In light of our study for the phase structure of PLSM at finite $T$, $\mu$, and $e B$, we can now speculate about the three-dimentional QCD phase-diagram. This was already analyzed in great details in Ref. \cite{Tawfik:2016ihn,Tawfik:2017cdx}, where the influence of finite magnetic field on the QCD phase-diagram, i.e. temperature vs. baryon chemical potential, was analyzed. Interested readers are kindly advised to consult Refs. \cite{Tawfik:2016ihn,Tawfik:2017cdx}, where  $T$-$eB$, $\mu$-$eB$, and $T$-$\mu$ QCD phase-diagrams are drawn, separately. The three-dimensional phase-diagram was depicted, as well \cite{Tawfik:2016ihn,Tawfik:2017cdx}. 

\begin{figure}[htb]
\centering{
\includegraphics[width=7.cm,angle=-90]{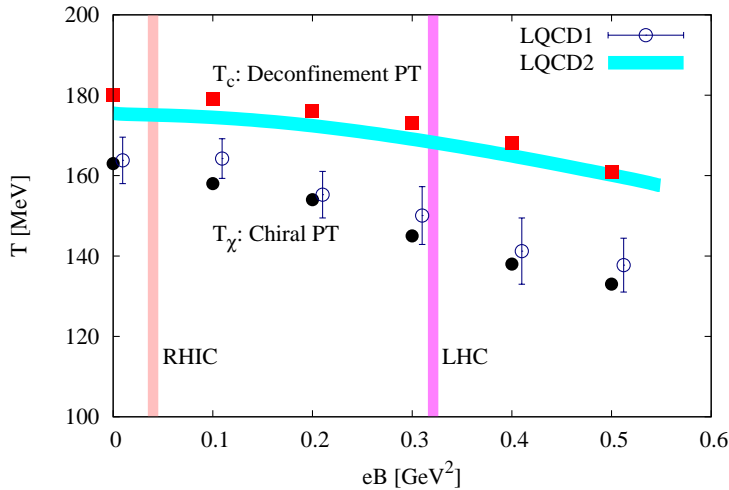}
\caption{(Color online)  The dependence of critical chiral and the critical deconfinement temperatures on finite magnetic field as determined from PLSM (closed symbols). The results are compared to recent $2+1$ lattice QCD simulations [curve (deconfinement) and open symbols (chiral)]  \cite{Refff25,Bali:2014Lattice-eB}. The vertical bands refer to averaged magnetic fields estimated at RHIC and LHC energies.
\label{fig:newfig1}}}
\end{figure}

In the present work, we concentrate the discussion on a detailed version of the $T$-$eB$ phase-diagram, Fig. \ref{fig:newfig1}. We distinguish between the critical chiral and the critical deconfinement temperatures. Also, we confront our calculations to recent  $2+1$ lattice QCD results \cite{Refff25,Bali:2014Lattice-eB}  and observe an excellent agreement.

Figure \ref{fig:newfig1} illustrates our PLSM estimations for both critical chiral and critical deconfinement temperatures, $T_{\chi}$ and $T_c$, respectively, in dependence on finite magnetic field. Both chiral (solid circles) and deconfinement (solid rectangles) calculations are confronted to recent $2+1$ lattice QCD simulations \cite{Refff25,Bali:2014Lattice-eB}. The open symbols stand for the lattice estimations for $T_{\chi}$, which are determined at the inflection point of the normalized entropy labeled as LQCD1 \cite{Bali:2014Lattice-eB}. The curved band represents the lattice extractions of the critical deconfinement temperature ($T_c$) at the inflection point of the strange quark-number susceptibility labeled as LQCD2 \cite{Refff25}.  

We recall that there are - at least - two different methods can be utilized in order to determine the critical temperatures. The first one, $T_{\chi}$, is based on the thermodynamic quantity $s/T^3$, where $s$ is the entropy density. The second one determines $T_c$ as positioned where the strange quark-number susceptibility reaches a maximum. It is apparent that the PLSM results in an excellent agreement with both lattice QCD predictions \cite{Bali:2014Lattice-eB,Refff25}. In light of this study, we conclude that both PLSM and lattice QCD simulations clearly indicate that the two types of critical temperatures (chiral and deconfindement) decrease with increasing $e B$, i.e. inverse magnetic catalysis.

Now a few remarks on the magnetic catalysis is in order. In Eq. (\ref{potential}), the first term, $ U(\sigma_l, \sigma_s)$, refers to the contributions of the valence quarks, the second term, $\mathbf{\mathcal{U}}(\phi, \phi^*, T)$] gives the gluonic potential contribution, while the last term [$\Omega_{\bar{q}q} (T, \mu _f, B)$] represents the contributions of the sea quarks. It is well-known that the physical mechanism for the magnetic catalysis relies on a competition between valance and sea quarks \cite{Fraga:2014-eB, Bruckmann:2013A-eBA,Bruckmann:2013A-eBB}. The mesonic potential has a remarkable effect at very low temperature \cite{Tawfik:2014uka}.  In our calculations, we noticed that the small value of $U(\sigma_l,\sigma_s)$ leads to an increase in the contributions of the sea quarks. Thus, applying finite magnetic field results in a suppression in the chiral condensates relevant to the restoration of the chiral symmetry breaking. This would explain the {\it ''inverse magnetic catalysis''}, which is characterized by a decrease in $T_c$ with increasing magnetic field. As illustrated in Fig. \ref{fig:newfig1}, where the critical temperatures of both types of the phase transitions decrease with increasing magnetic field. It should be noted that this phenomenon is also observed in the lattice QCD simulations \cite{Refff25,Bali:2014Lattice-eB}.

It is noteworthy recalling that Ref. \cite{Refff26} reported on opposite results, i.e. on {\it direct} magnetic catalysis. The first term in the free energy, Eq. (\ref{potential}), $U(\sigma_l, \sigma_s)$, refers the contribution of the valance quarks. In a previous work \cite{Tawfik:2014uka}, we have analyzed the temperature dependence of this specific potential at vanishing chemical potential. We have found that this has a small effect at high temperatures. But it plays an important role in assuring spontaneous chiral symmetry breaking. The PLSM describing the quark-hadron phase structure, where the valence and sea quarks are simultaneously implemented, Eq. (\ref{potential}). We propose that the physical mechanism of the magnetic catalysis relies on an interplay between the contributions of sea and valance quarks \cite{Bruckmann:2013A-eBB, Fraga:2014-eB}. We have observed that the magnetic field increasingly suppresses the chiral condensates. Thus, the restoration of the chiral symmetry breaking takes place at lower temperatures. Although, this is unlikely that alone the number of the quark flavors is capable to explain when to obtain {\it inverse} or {\it usual} magnetic catalysis, we emphasize that our calculations assume $2+1$ quarks flavors. For the sake of completeness, we intend in a future work to investigate the magnetic catalysis in SU($4$) PLSM, as well.

%--------------------------------------------------------------------------------
%                                        Maseom states
%--------------------------------------------------------------------------------

\subsection{Chiral phase-structure of various meson states in finite magnetic field} 
\label{mesons}

In a previous work, co-authors AT and AD have studied {\it sixteen} meson states in thermal and dense medium from SU($3$) PLSM \cite{Tawfik:2014gga}, but at vanishing magnetic field. At finite magnetic field, an estimation for the magnetic effects on {\it four} meson states was presented in Ref. \cite{Tawfik:2014hwa}. In order to conduct a systematic study for the effects of the magnetic field strength on nonet {\it sixteen} meson states, the chiral phase-structure of each meson-state shall be analyzed in dependence on temperature, baryon chemical potential and finite magnetic field. 
\begin{itemize}
\item We start with an approximate range for the magnetic field strength that may be created in heavy-ion collisions at RHIC and LHC energies \cite{Skokov:2009, Elec:MagnetA, Elec:MagnetB}, $eB=1-25\;m^2_{\pi}$.
\item Then, we construct the temperature dependence of all meson states corresponding to the lowest Matsubara frequencies. 
\item Last but not least, the chiral phase-structure of the meson states are then given in a wide range of baryon chemical potentials. 
\end{itemize}

In quantum field theory, the hadron mass can be determined from the second derivative of the equation of motion relative to the hadron field of interest. The free energy $\mathcal{F} (T, \, \mu,\, eB, \beta)$ apparently encodes details about the equation of motion, where $\beta$ represents the corresponding meson field. Assuming that the contribution of the quark-antiquark potential to the Lagrangian vanishes in the vacuum, the meson potential determines the mass matrix can be given as
\bea
m_{i,ab}^2 &=& \left. \frac{\partial^2 \mathcal{F} (T, \, \mu,\, eB, \beta) }{\partial \, \beta_{i,a} \, \partial \, \beta_{i,b} }\right|_{\rm min}, \label{vaccum}
\eea
where $i$ stands for (pseudo)scalar and (axial)vector mesons and $a$ and $b$ are integers ranging between $0$ and $8$. Equation (\ref{vaccum}) can be split into two different terms.
\begin{itemize}
\item The first one is related to the vacuum, where the meson masses are developed from the nonstrange ($\sigma_l$) and strange ($\sigma _s$) sigma fields. More details about the estimation of the masses in vacuum are given in appendix \ref{app:Mass}.  In this part, the effect of the magnetic field requires numerical estimations for nonstrange and strange sigma-fields. 
\item The in-medium term, in which the magnetic field  is included reads
\bea
m_{i,ab}^2  &=& \left. \nu_{c}\sum_{f} \frac{|q_f| B \, T}{(2 \pi)^2} \,  \sum_{\nu = 0}^{\infty}  (2-\delta _{0 \nu })    \int_0^{\infty} dp_z  \right.  \frac{1}{2E_{B, f}} \biggl[ (n_{q,f,B} + n_{\bar{q},f,B} ) \biggl( m^{2}_{f,a b} - \frac{m^{2}_{f,a}
 m^{2}_{f, b}}{2 E_{B, f}^{2}} \biggr) 
\nonumber   \\  && \hspace*{6.5cm} +(b_{q,f,B} + b_{\bar{q},f,B}) \biggl(\frac{m^{2}_{f,a}  m^{2}_{f, b}}{2 E_{B, f}\; T}
\biggr) \biggr],  \label{eq:ftmass}
\eea 
where $\nu_c=2\, N_c$. Apparently, this expression gives the meson-mass modifications which can be estimated from PLSM, Eq. (\ref{vaccum}), and the diagonalization of  the resulting quark-mass matrix, as well.  Details about other quantities in Eq. (\ref{eq:ftmass}) can be taken from Ref. \cite{V. Tiwari:2009,Tawfik:2014gga}:
\begin{itemize}
\item The quark mass derivative with respect to the meson fields ($\lambda_{i,a}$);   
$ m^2_{f,a} \equiv \partial m^2_f/\partial \lambda_{i,a}. $
\item The quark mass with respect to meson fields ($\lambda_{i,a} \partial  \lambda_{i,b}$), 
$ m^2_{f,{ab}} \equiv \partial m^2_f/\partial \lambda_{i,a} \partial  \lambda_{i,b}$.
\item Correspondingly, the antiquark function $b_{\bar{q},f,B}(T,\,eB,\,\mu_f) = b_{q,f,B}(T,\, eB,\,-\mu_f)$, where
\bea
b_{q,f,B}(T,\,eB,\,\mu_f)= n_{q,f,B}(T,\,eB,\,\mu_f) (1-n_{q,f,B}(T,\,eB,\,\mu_f)).
\eea
\item The normalization factors for quark ($b_{q,f,B}=3(n_{{q},f} )^2 - c_{q,f,B}$) and that for antiquark read $b_{\bar{q},f,B}=3(n_{{\bar{q}},f,B} )^2 - c_{{{\bar{q}}},f,B}$,  
\bea
n_{q,f,B} &=& \frac{\Phi e^{-\,E_{q,f}/T} + 2 \Phi^* e^{-2\,E_{q,f}/T} + e^{-3\,E_{q,f}/T}}{1+3(\phi+\phi^* e^{-E_{q,f}/T}) e^{-E_{q,f}/T}+e^{-3E_{\bar{q},f}/T}}, \\
n_{\bar{q},f,B} &=& \frac{\Phi^* e^{-E_{\bar{q},f}/T} + 2 \Phi e^{-2E_{\bar{q},f}/T} + e^{-3E_{\bar{q},f}/T}}{ 1+3(\phi^*+\phi e^{-E_{\bar{q},f}/T}) e^{-E_{\bar{q},f}/T}+e^{-3E_{\bar{q},f}/T}},
\\ 
c_{{q},f,B} &=& \frac{\Phi e^{-\,E_{q,f}/T} +4 \Phi^* e^{-2\,E_{q,f}/T} +3 e^{-3\,E_{q,f}/T}}{1+3(\phi+\phi^* e^{-E_{q,f}/T})\, e^{-E_{q,f}/T}+e^{-3E_{\bar{q},f}/T}}, \\
c_{\bar{q},f,B} &=&  \frac{\Phi^* e^{-E_{\bar{q},f}/T} + 4 \Phi e^{-2E_{\bar{q},f}/T} +3 e^{-3E_{\bar{q},f}/T}}{ 1+3(\phi^*+\phi e^{-E_{\bar{q},f}/T})\, e^{-E_{\bar{q},f}/T}+e^{-3E_{\bar{q},f}/T}}.
\eea
\item The quark and antiquark dispersion relations, $E_{\bar{q},f}(T,\,eB,\,\mu)$ and $E_{q,f}(T,\,eB,\,-\mu)$, respectively are equivalent to $E_{B, f}$ in Eq. (\ref{eq:moddisp}).
\end{itemize}
\end{itemize}  

It should noticed that due to the mixing between the (pseudo)scalar and (axial)vector sector through the covariant derivative, the tree-level expressions of the pseudoscalars and some scalars aren't mass eigenstates. When the corresponding wave functions are renormalized to the constants $Z$, such a mixing can be resolved. Details can be found in Ref. \cite{Rischke:2012}.

\subsubsection{Dependence  of various meson states on temperature \label{sec:Tdepnd}}

\begin{figure}[htb]
\centering{
\includegraphics[width=3.9cm,angle=-90]{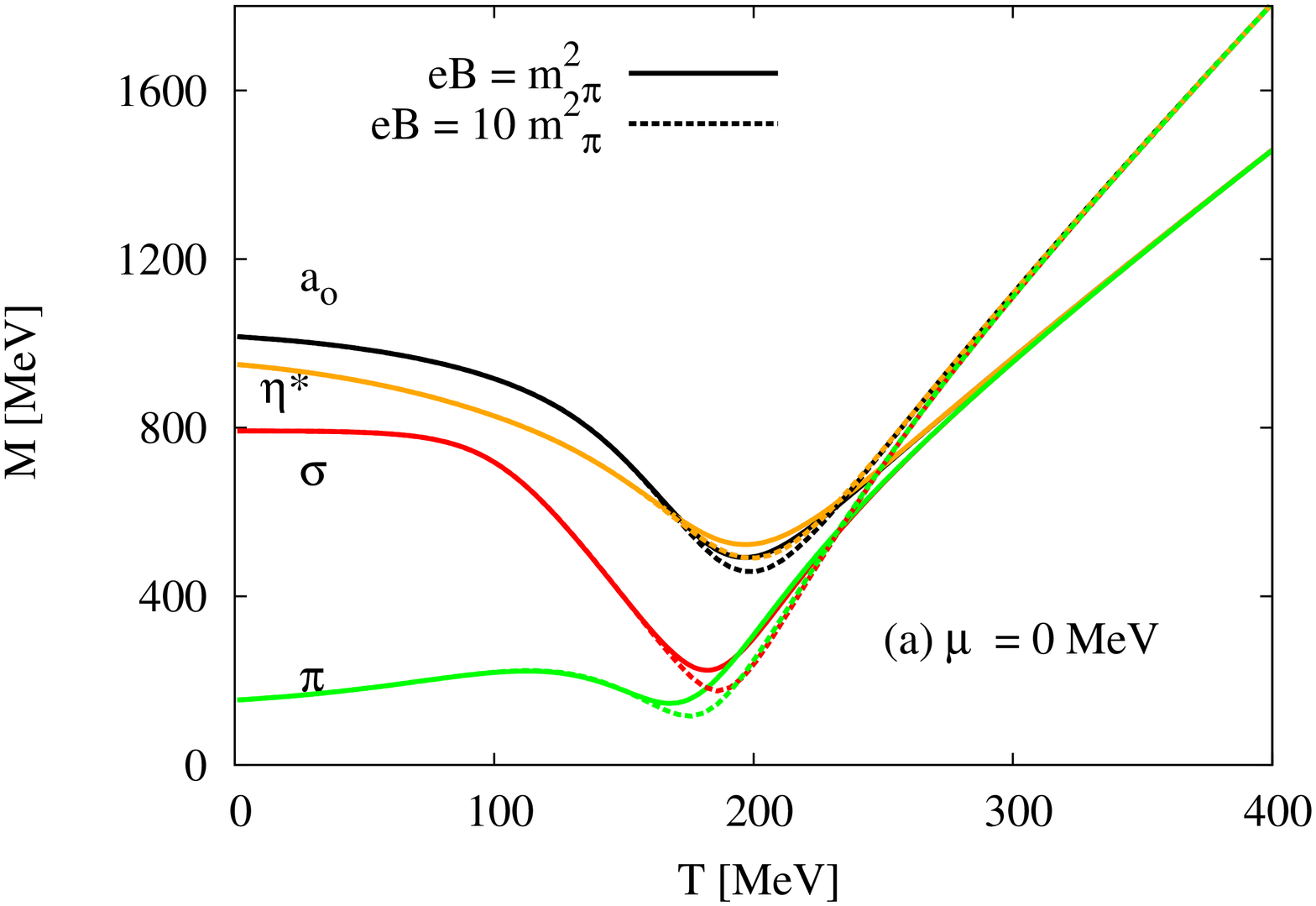}
\includegraphics[width=3.9cm,angle=-90]{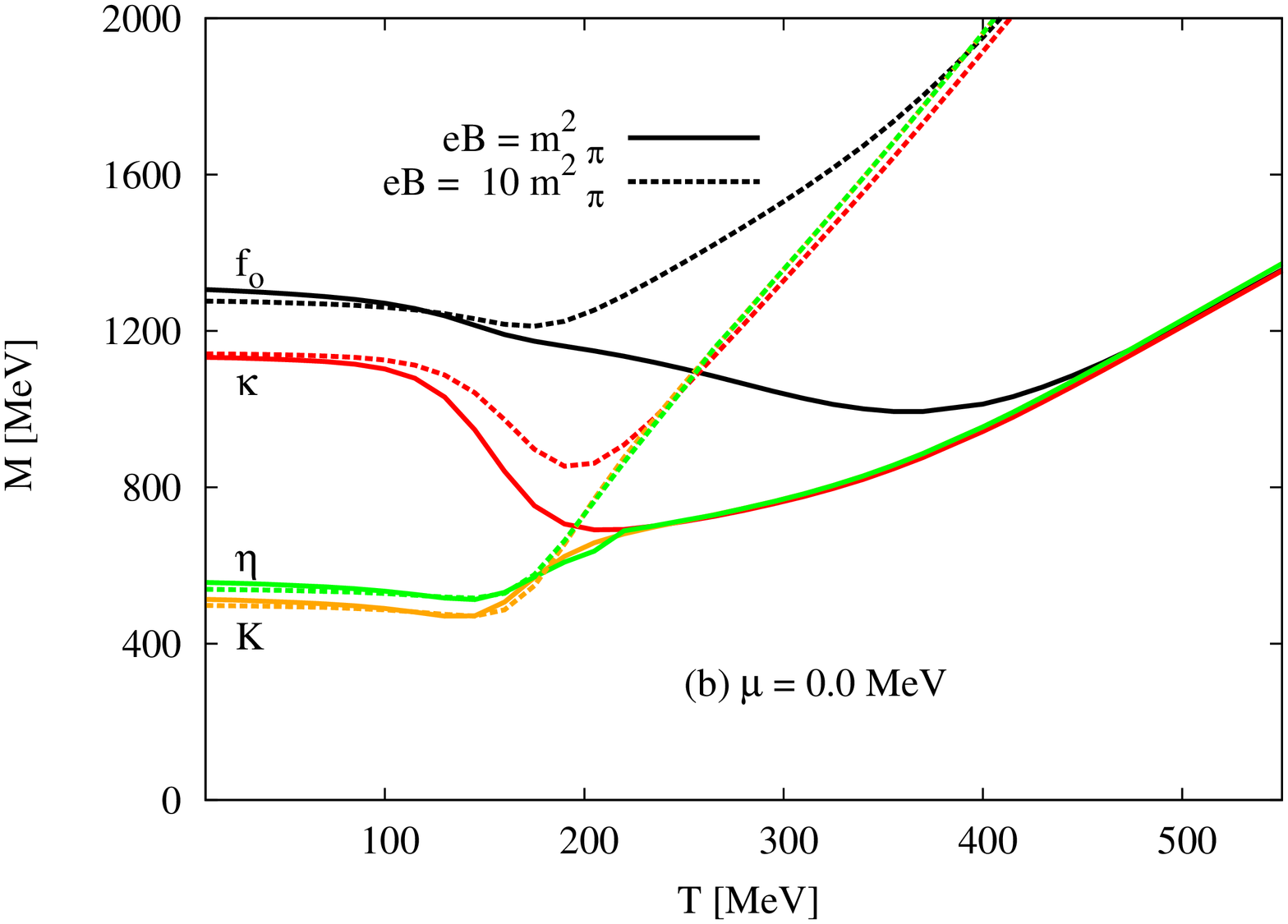}
\includegraphics[width=3.9cm,angle=-90]{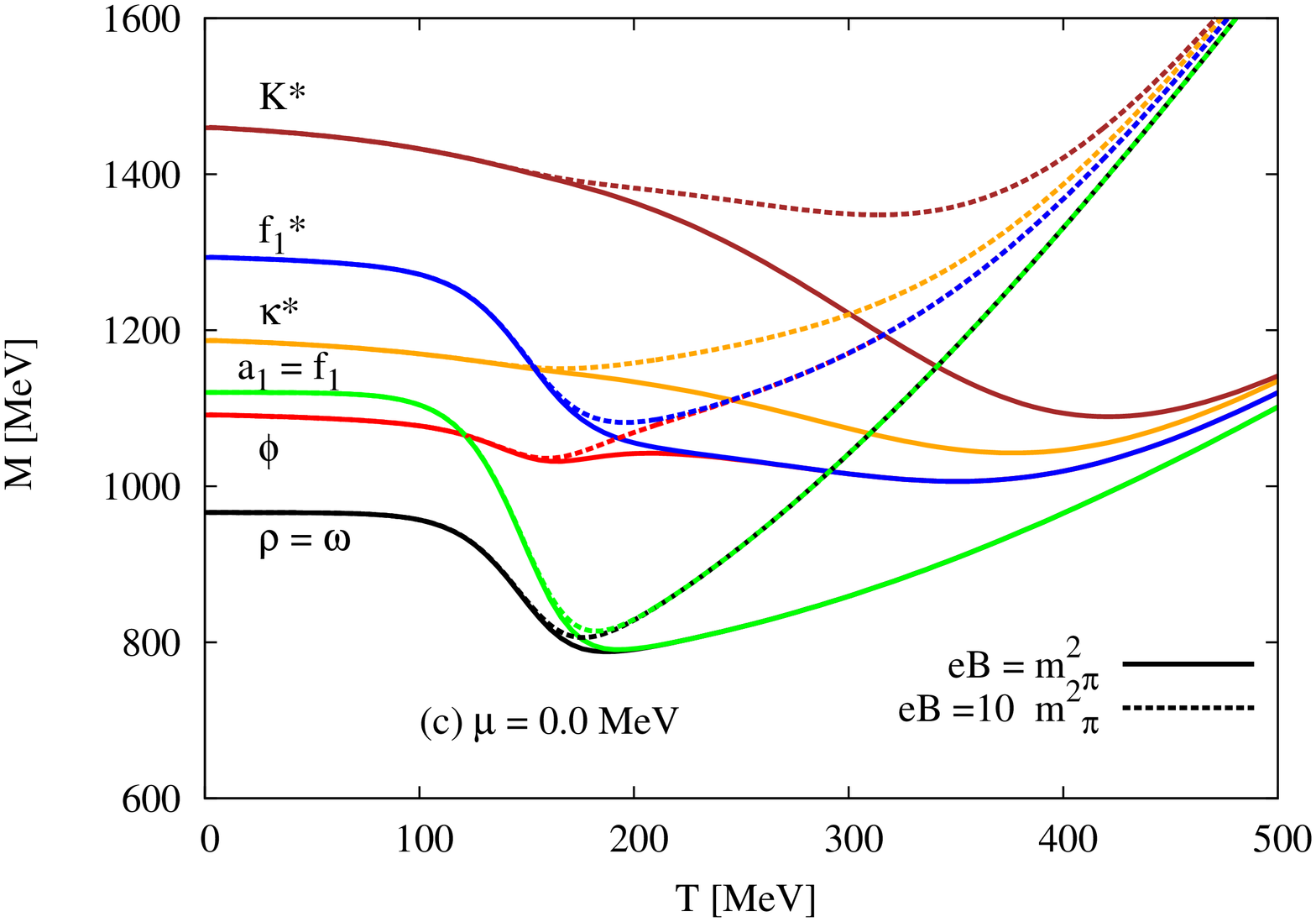}\\
\includegraphics[width=3.9cm,angle=-90]{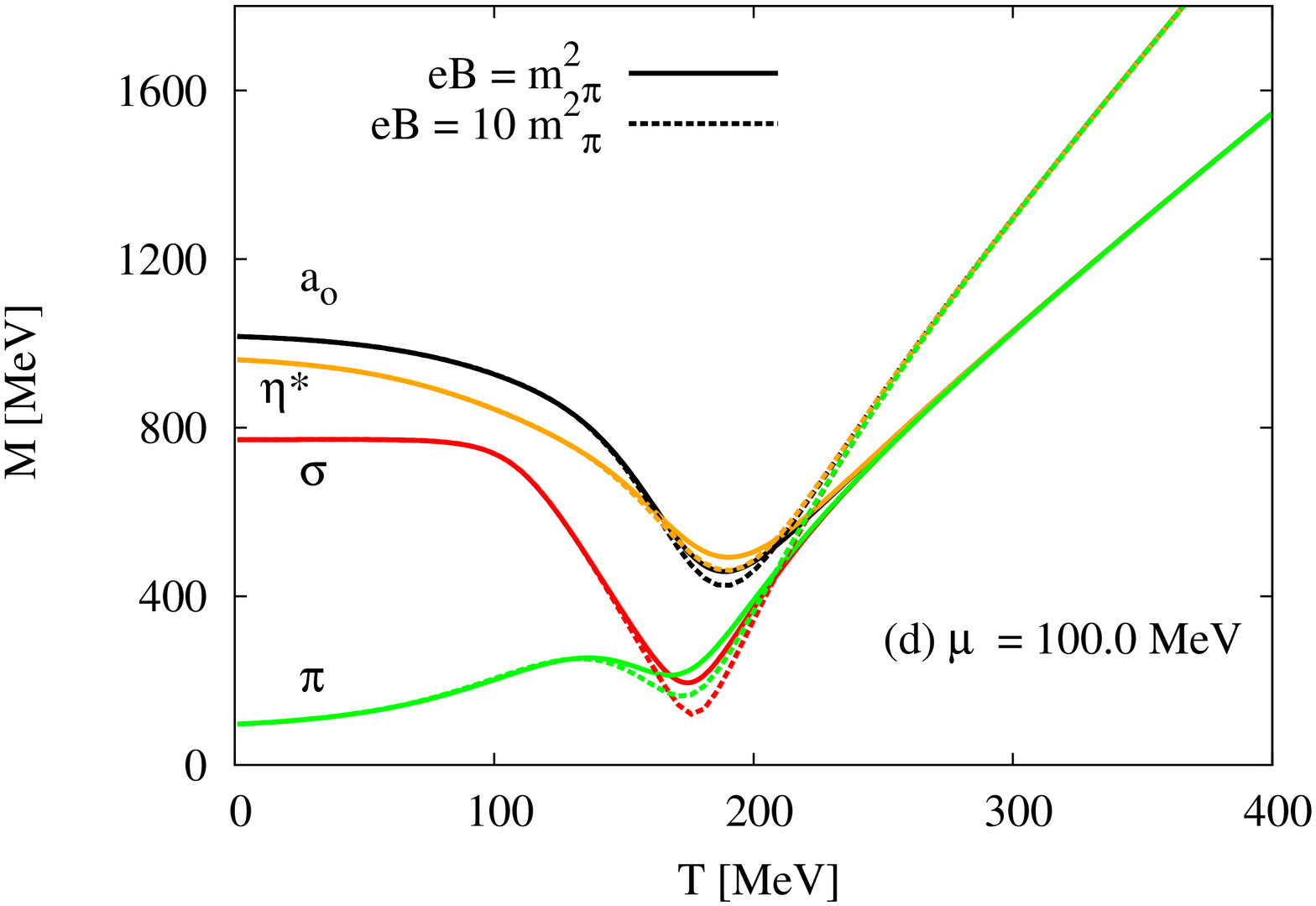}
\includegraphics[width=3.9cm,angle=-90]{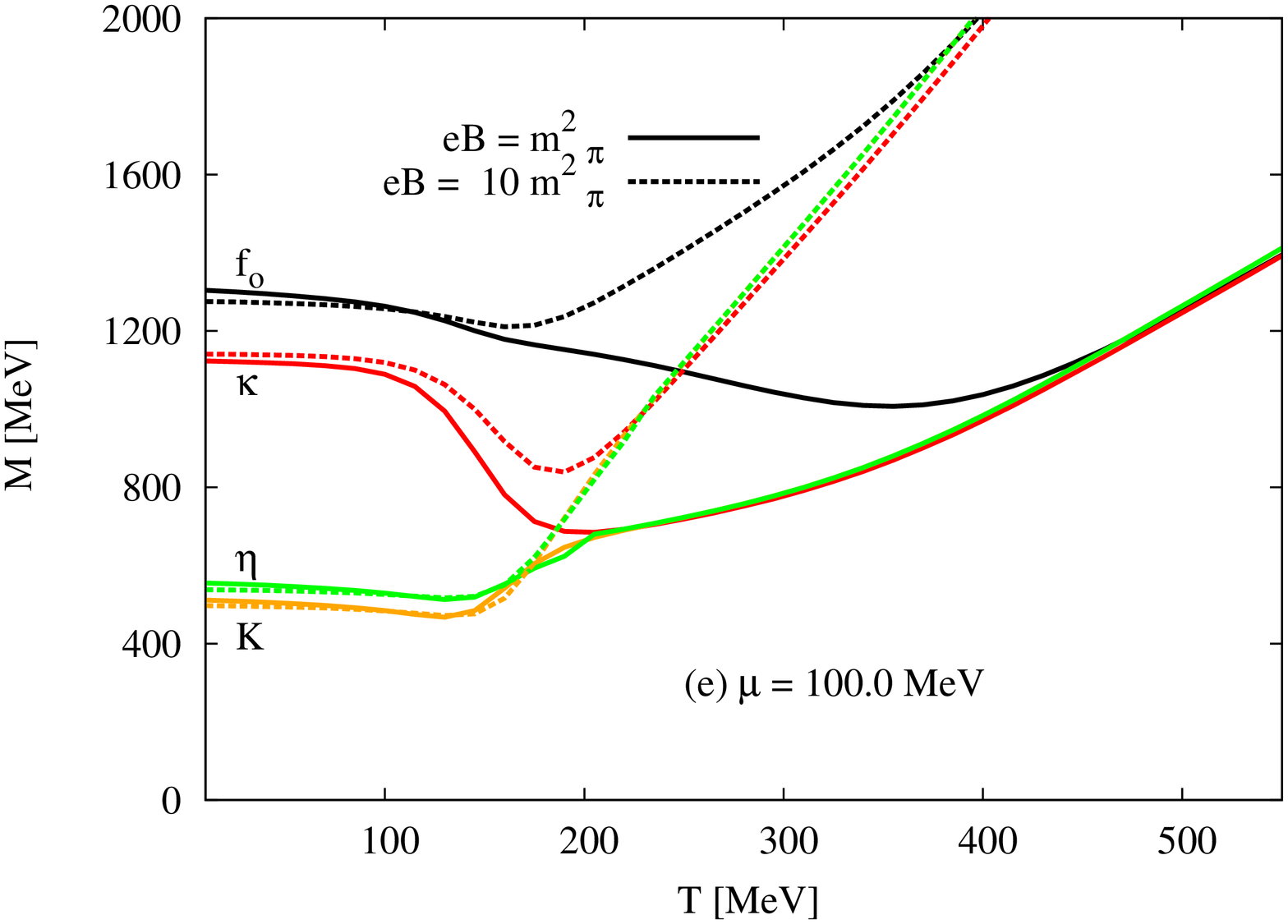}
\includegraphics[width=3.9cm,angle=-90]{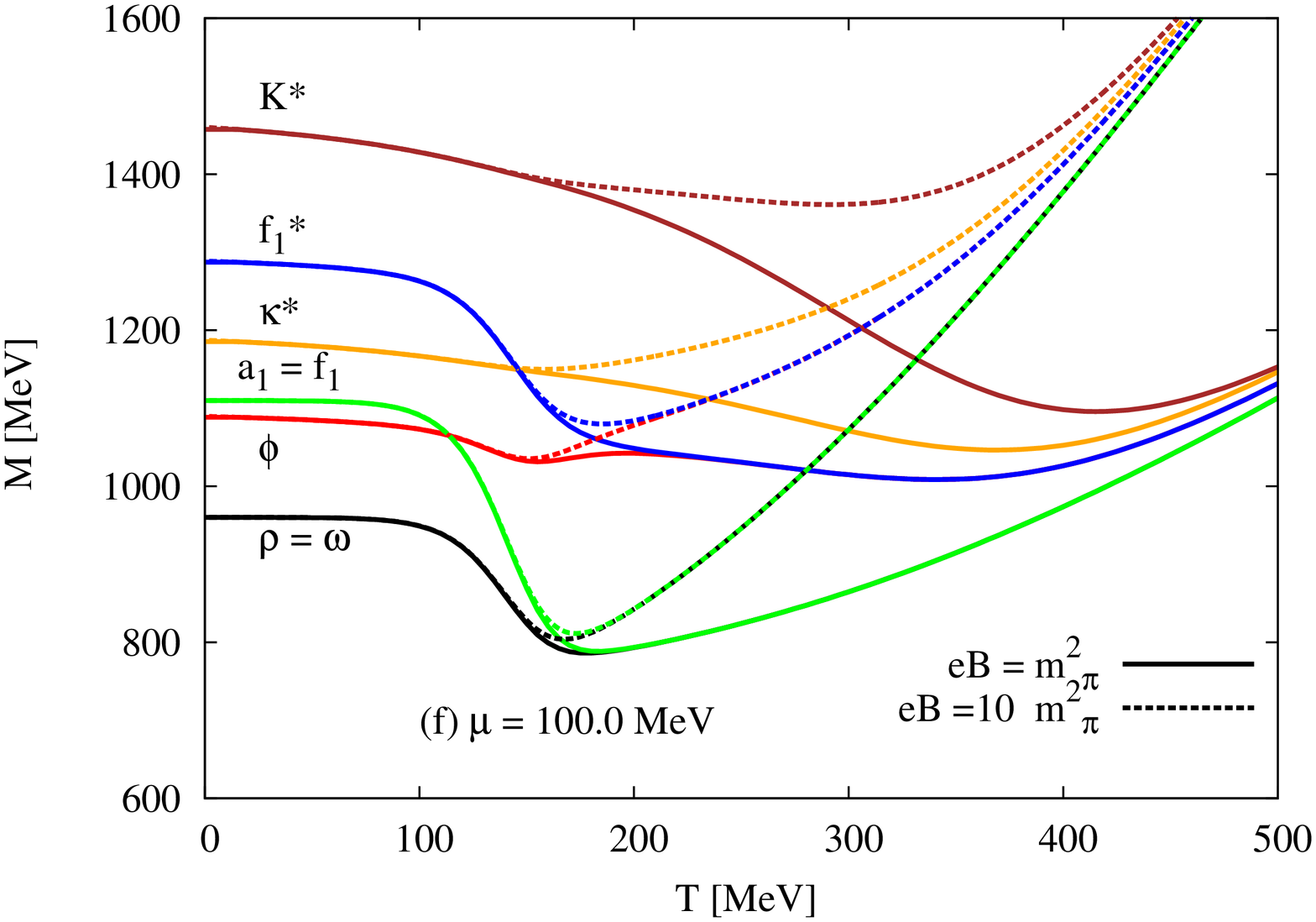}\\
\includegraphics[width=3.9cm,angle=-90]{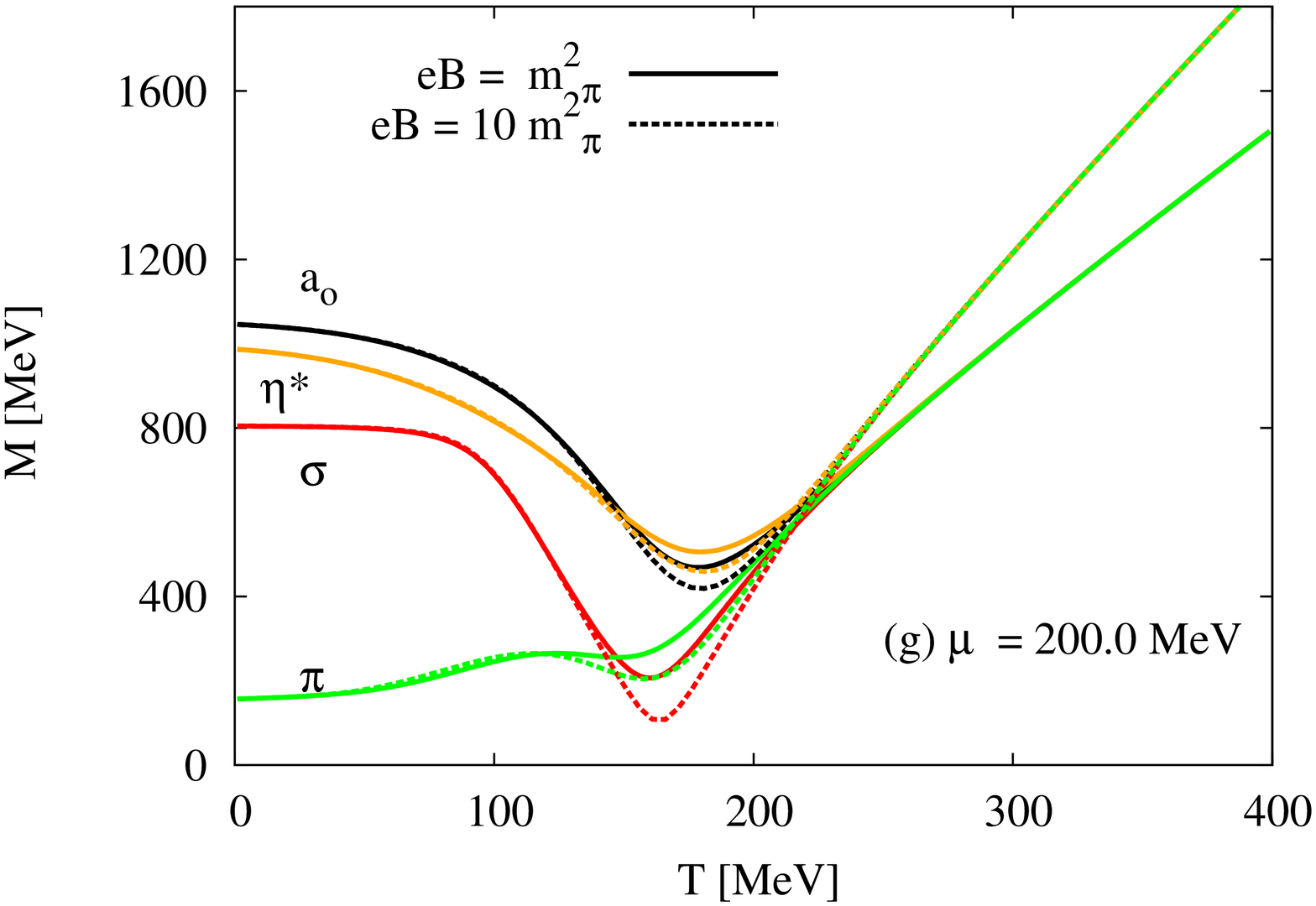}
\includegraphics[width=3.9cm,angle=-90]{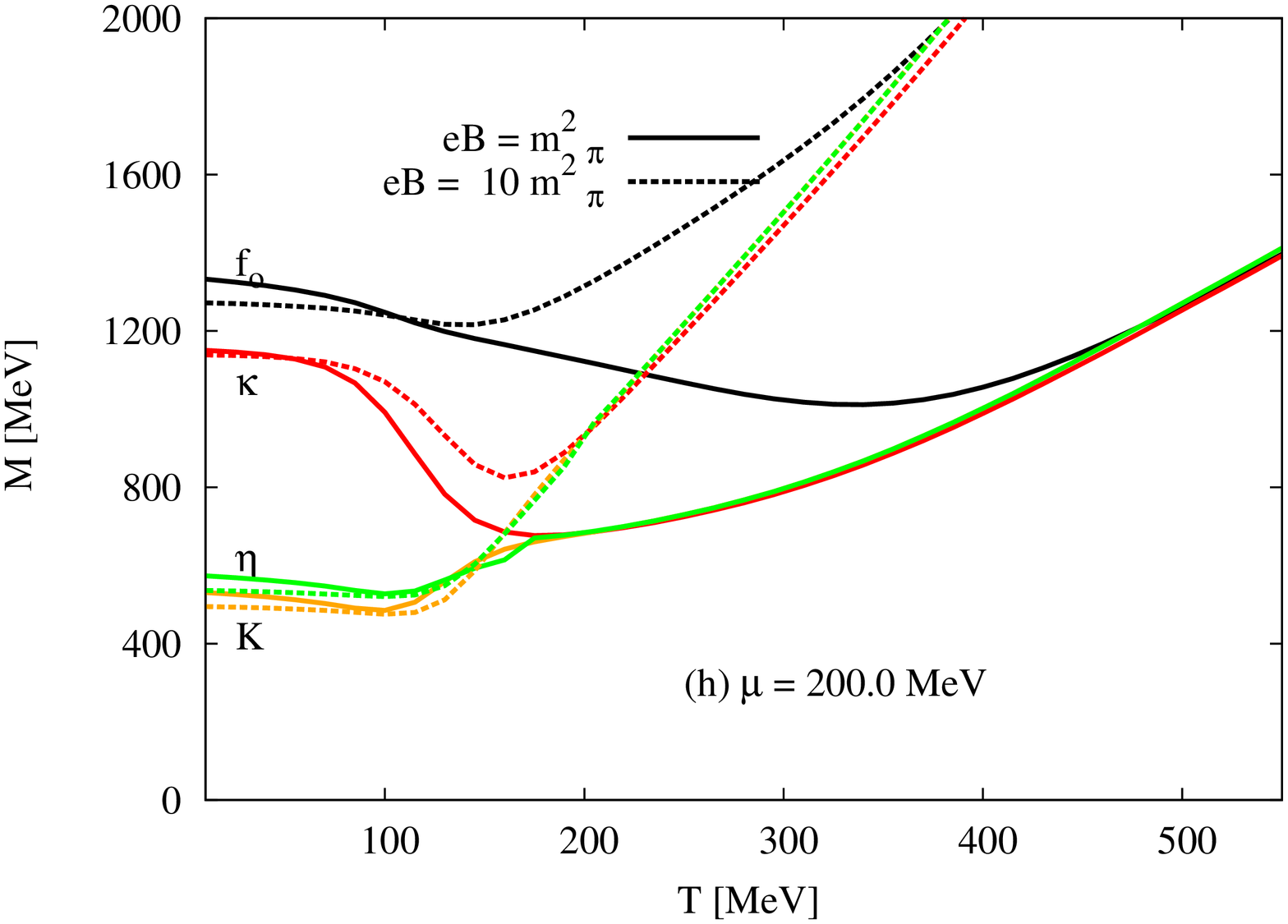}
\includegraphics[width=3.9cm,angle=-90]{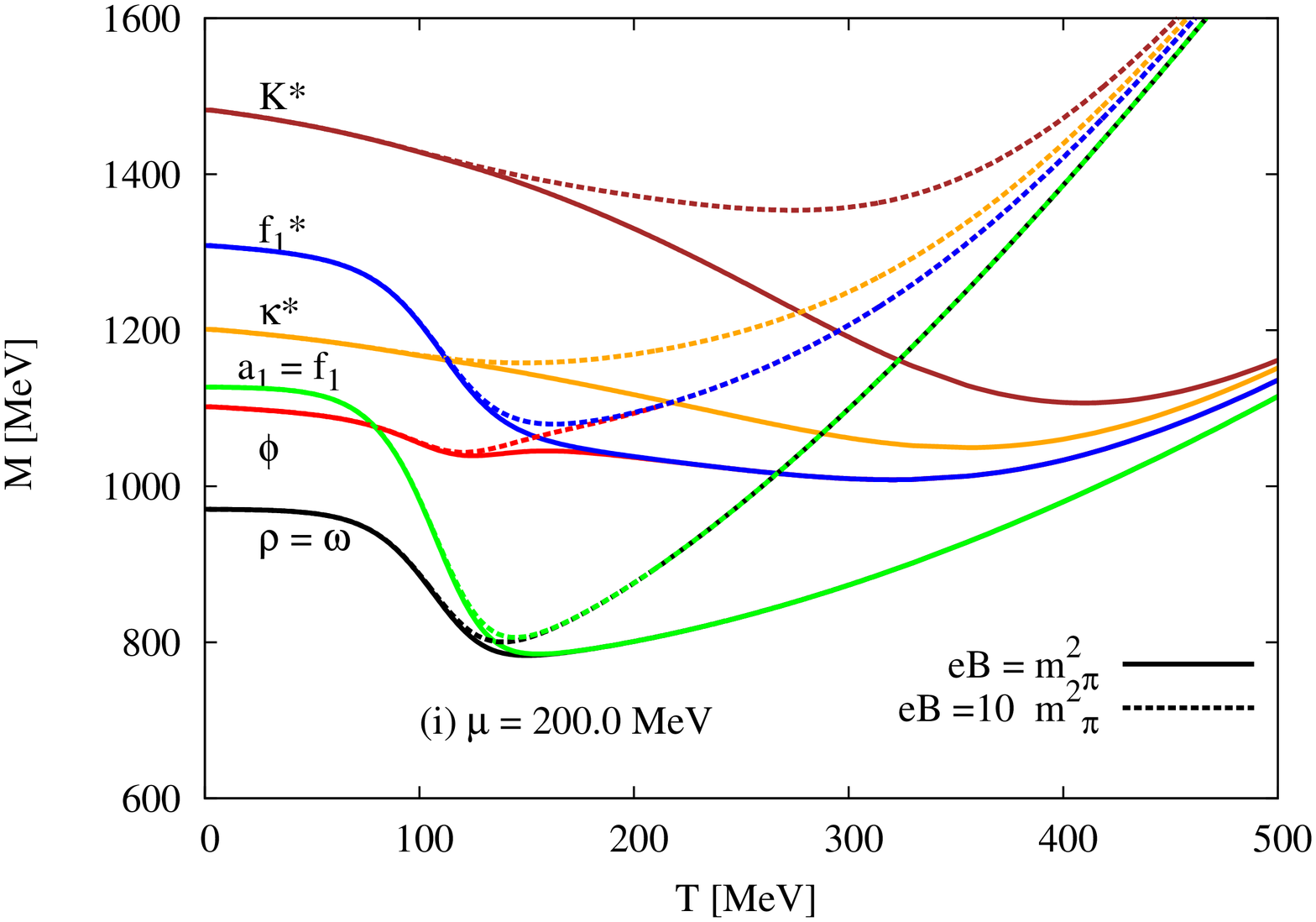}
\caption{\footnotesize (Color online) The temperature dependence of (pseduo)scalar and (axial)vector meson states calculated at finite magnetic fields; $eB=m_{\pi}^2$ (solid curves) and $eB=10 m_{\pi}^2$ (dotted curve) at different baryon chemical potentials; $\mu=0~$MeV (top panels), $\mu=100~$MeV (middle panels) and $\mu=200~$MeV (bottom panels) is depicted. \label{fig:Tmass}
}}
\end{figure}

Figure \ref{fig:Tmass} shows the (pseudo)scalar and (axial)vector meson states with $U(1)_A$-anomaly as functions of temperatures at fixed magnetic field strengths and fixed baryon chemical potentials. The left-hand panels give $\left[\mathrm{a}_0(980),\,\mathrm{\sigma}(800)\right]$ scalar and $\left[\mathrm{\eta}^*(957),\, \mathrm{\pi}(134) \right]$ pseudoscalar states, while the scalar $\left[f_0(1200),\, \kappa(1425) \right]$ and pseudoscalar $\left[\mathrm{\eta}(547), \mathrm{K}(497)\right]$ meson states are presented in the middle panels. The right-hand panels illustrate the vector $\left[\mathrm{\rho}(775),\, \mathrm{K}^*(891),\,\mathrm{\omega}(782),\,\mathrm{\phi}(1019) \right]$ and axialvector $\left[\mathrm{a}_1(1030),\,\mathrm{f}_1(1281),\, \mathrm{K}_1^*(1270),\, \mathrm{f}_1^*(1420)\right]$ meson states. The solid curves represent the results at $eB=m_{\pi}^2$. The results at $\mu=0.0~$MeV are depicted in the top panels. Middle and bottom panels are devoted to the calculations at $\mu=100~$MeV and $\mu=200~$MeV, respectively. The results at  $eB=m_{\pi}^2$ are given as dashed curves. To distinguish between such curves, one should notice that both curves corresponding to one meson state are mostly bundled at low temperatures. At very high temperatures, many - if not all - curves approach some asymptotic limits. In left-hand panels, we observe that increasing magnetic field slightly reduces the meson masses, especially at the critical temperatures. In middle and right-hand panels, the opposite is obtained with one exception for Kaons. This might be a subject of experimental justification, especially by mean of future facilities such as the Facility for Antiproton and Ion Research (FAIR) and the Nuclotron-based Ion Collider fAсility (NICA).

The mass gap between the various meson states can be estimated from the bosonic thermal contributions. The masses of bosons are given as functions of the pure sigma fields; the nonstrange ($\sigma_l$) and the strange ($\sigma_s$). At low temperatures, the bosonic thermal contributions are dominant and become stable and finite relative to the corresponding vacuum value of the meson state until they reach the chiral temperature ($T_{\chi}$) related to the given meson state. At high temperatures, the fermionic (quark) thermal contributions complete the thermal behavior of these states. As they increased with the temperature, this leads to degenerate meson masses. It is obvious that the effects of the fermionic (quark) thermal contributions are negligible at low temperatures. 

As depicted in Fig. \ref{fig:Tmass}, the effects of finite magnetic field on the thermal contributions of the meson states can be divided into three phases. 
\begin{itemize}
\item The first region is dominated by the bosonic thermal contributions which remain ineffective until the second region takes place.
\item The second region, beyond the chiral phase-transition of the given meson state, here the influence of the magnetic field becomes obvious. We observe that, increasing the magnetic field accelerates the chiral phase-transition of the given meson. With accelerating we mean that the phase transition takes place at lower temperatures or chemical potentials. As a result, the chiral critical temperature decreases with increasing magnetic field strength.  
\item The last (third) region represents the fermionic thermal contributions. 
\end{itemize}
We conclude that the magnetic field has an evident effect on the quarks and apparently accelerates and sharpens the quark-hadron phase transition.

We observe that at temperatures exceeding the critical values (corresponding to each meson state), the meson masses become degenerate. This can be understood due to the effect of the fermionic fluctuations on the chiral symmetry restoration \cite{Schaefer:2009}, especially on the strange condensate ($\sigma_s$). Such an effect seems to melt the nonstrange condensate ($\sigma_l$) in a way faster than that of the nonstrange one ($\sigma_l$), Fig. \ref{fig:sbtrc1}. At very high temperatures, the mass gap between mesons seems to disappear and to decrease due to melting strange condensate ($\sigma_s$). This mass gap appears again at low temperatures, where the nonstrange condensate remains finite. At temperatures higher than the critical one, only strange condensate remains finite. This thermal effects is strongly related to the degeneration of the meson masses.

Also, Fig. \ref{fig:Tmass} shows that the $\mathrm{\sigma}$, $\mathrm{\pi}$, $\mathrm{a}_0$, and $\mathrm{\eta}^*$ meson states become degenerated through a first-order phase transition and apparently keep this at higher temperatures assuring a completion of the chiral symmetry restoration. The middle panels depict $\mathrm{\kappa}$ meson state and illustrates that it drops to the $\mathrm{\eta}$ and $\mathrm{K}$ meson states through first-order phase transition. $\mathrm{f}_0$ is degenerated at a higher temperature as well through first-order phase transition. In right-hand panels, the meson states $\mathrm{f}_1^*$, $\mathrm{\phi}$, $\mathrm{a}_1$, $\mathrm{f}_1$, $\mathrm{\rho}$, and $\mathrm{\omega}$ become degenerated through chiral phase-transitions, whereas $\mathrm{K}^*$ and $\mathrm{\kappa}^*$ become coincident at very high temperature. 

Increasing the baryon chemical potential influences the behavior of the thermal contributions of the different meson states, as well. We find that the increase in the baryon chemical potential (from top to bottom panels) enhances the degeneration of the various meson masses. For example, $\mathrm{\sigma}$ and $\mathrm{\pi}$ states degenerate at the chiral temperature $T_{\chi}\sim 191.5~$MeV at $\mu=0~$MeV. But at $\mu=100~$MeV, the critical temperature drops to $\sim 186.5~$MeV and $\mu=0~$MeV, it decreases to $\sim 178.7~$MeV. Thus, we conclude that the chiral temperature associated to the different meson states decreases with increasing the baryon chemical potential. This refers to a remarkable in-medium effect that can be verified by the future facilities running high-density heavy-ion collision experiments.

Our goal is not only the characterization of the meson spectra in thermal and dense medium but also the description of their vacuum phenomenology in finite magnetic filed. In Tab. \ref{masscomp}, an extensive comparison is summarized between the results on different scalar and vector meson (nonets) from various effective models; PLSM (present work) and PNJL \cite{NJL:2013,P. Costa:PNJLA,P. Costa:PNJLB,P. Costa:PNJLC,Blanquier} and from the very recent compilation of PDG \cite{PDG:2018}, the lattice QCD calculations \cite{HotQCD,PACS-CS} and QMD/UrQMD simulations \cite{URQM}. We conclude that our PLSM-calculations are remarkably precise, especially for some light mesons at a vanishing temperature. They are very comparable with measurements and lattice calculations, as shall be elaborated in the following section.

Only pseudoscalar and vector meson sectors are available from HotQCD \cite{HotQCD} and PACSCS collaborations \cite{PACS-CS}. It is obvious that our estimations for the meson masses agree well with previous calculations \cite{Schaefer:2009,V. Tiwari:2009,Rischke:2012,Tawfik:2014gga} expect for mixing strange with nonstrange scalar states, where it was reported that one gets various states below $1~$GeV and another one above $1$ GeV \cite{Rischke:2012}. The agreement between PLSM- and the available lattice-calculations is remarkably excellent.

%{\fontsize{1}{1}\selectfont
%\hspace{-10.0cm} 
\begin{table}[htb]
\smaller
\begin{tabular}{ |c | c | c | c | c | c | c | c ||}
\hline
& Sector & \begin{tabular}{c} Meson \\States \end{tabular}  & PDG \cite{PDG:2018} & \begin{tabular}{c}UrQMD\\ Model \cite{URQM}\end{tabular}  & \begin{tabular}{c}Present\\ Work\end{tabular} &\begin{tabular}{c} PNJL\\ Model \cite{NJL:2013,P. Costa:PNJLA,P. Costa:PNJLB,P. Costa:PNJLC,Blanquier} \end{tabular} &\begin{tabular}{c|c}\multicolumn{2}{ c }{Lattice QCD Calculations} \\
\hline
Hot QCD\cite{HotQCD} & PACS-CS  \cite{PACS-CS} \\\end{tabular} \\
\hline \hline
&\begin{tabular}{c}
Scalar \\$J^{PC}=0^{++}$
\end{tabular}
&\begin{tabular}{c}
$\mathrm{a}_{0}$~\\ ~$\mathrm{\kappa} $~\\~$\mathrm{\sigma}$~\\~$ \mathrm{f}_{0}$~\\
\end{tabular} 
&\begin{tabular}{c}
$\mathrm{a}_{0}(980 {\pm 20})$~\\$\mathrm{K}_{0}^* (1425 {\pm 50})$\\ \\$ \mathrm{f}_{0}(1281.0 \pm 0.5)$~\\
\end{tabular} 
&\begin{tabular}{c}
$984.7$~\\ $1429$ \\$400 - 1200$\\$ 1200-1500$~\\
\end{tabular}  
&\begin{tabular}{c}
$1015.73  $~\\ ~$ 1115 $~\\~$ 800 $~\\~$ 1102.8$~\\
\end{tabular} 
&\begin{tabular}{c}$837 $~\\ ~$ 1013 $~\\~$ 700 $~\\~$ 1169$~\\ \end{tabular} & \\  
\hline
&\begin{tabular}{c}\\
Pseudoscalar \\
$J^{PC}=0^{-+}$
\end{tabular} 
&\begin{tabular}{c}
$\mathrm{\pi}$~\\ ~$\mathrm{K} $~\\~$\mathrm{\eta}$~\\~$\mathrm{\eta}^{'}$~\\
\end{tabular} 
&\begin{tabular}{c}
$\mathrm{\pi}^0 (134.9770 {\pm 0.0005} )$~\\ $\mathrm{K}^0$ ($497.611 {\pm 0.013}$) \\ $\mathrm{\eta}(547.862 {\pm 0.17} )$~\\$\mathrm{\eta}^{'}(957.78 { \pm 0.06})$\\
\end{tabular}  
&\begin{tabular}{c}
$139.57$~\\ ~$493.68$~\\~ $517.85$~\\~$\,957.78$~\\
\end{tabular} 
&\begin{tabular}{c}
$150.4$~\\ ~$\,509$~\\~ $ 553 $~\\~$ 949.7 $~\\
\end{tabular}  &\begin{tabular}{c}$126$~\\ ~$\, 490$ \\ $505$ \\ $949$\\ \end{tabular} &
\begin{tabular}{l}$134 {\pm 6} ~~~~~~~~~~135.4 {\pm 6.2}$ \\ $422.6 {\pm 11.3}\, \,~~~498 {\pm 22} $\\ $579 {\pm 7.3}\, \,~~~~~~~688 {\pm 32} $\\$\, \,~~~~~~~~~ $~\\ \end{tabular}
 \\  
\hline 
&\begin{tabular}{c}
Vector\\ $J^{PC}=1^{--}$
\end{tabular}
&\begin{tabular}{c}
~$\mathrm{\rho}$~\\ ~$\mathrm{\omega}_X$~\\ ~ $\mathrm{K}^*$ ~\\ ~$\mathrm{\omega}_y$~\\
\end{tabular}
 &\begin{tabular}{c}
$\mathrm{\rho}(775.26 {\pm 0.25})$\\ ~$\mathrm{\omega} (782.65 {\pm 0.12})$~\\ ~ $\mathrm{K}^* (891.76 {\pm 0.25})$ ~\\ ~$\mathrm{\phi}(1019.461 {\pm 0.016})$~\\
\end{tabular} 
&\begin{tabular}{c}
~$771.1$~\\ ~$782.57$~\\ ~ $891.66$ ~\\ ~$ 1019.45$~\\
\end{tabular}
&\begin{tabular}{c}
~$745$~\\ ~$745$~\\~ $894$ ~\\ ~$ 1005$~\\
\end{tabular}
 &\begin{tabular}{c}$764.08$~\\ ~$764.08 $~\\~ $899.96$~\\~$1025.79$~\\ \end{tabular} &\begin{tabular}{l} $\,\; 756.2 {\pm 36}~~~~~~597 {\pm 86} $\\ $\; 884 {\pm 18}\, ~~~~~~~~861 {\pm 23} $\\ $\, 1005 {\pm 93}\, ~~~~~~~1010.2 {\pm 77} $\\$\, \,~~~~~~~~~ $~\\ \end{tabular}
 \\
\hline 
&\begin{tabular}{c}
Axial-Vector\\ 
$J^{PC}=1^{++}$
\end{tabular}
&\begin{tabular}{c}
~$\mathrm{a}_1$~\\ ~$\mathrm{f}_{1x}$~\\ ~ $\mathrm{K}_{1}^{*}$ ~\\ ~$\mathrm{f}_{1y}$~\\
\end{tabular}
 &\begin{tabular}{c}
$a_1 (1230 \pm 40)$\\ ~$f_{1}(1281.9 { \pm 0.5})$~\\ ~ $K_{1}^{*} (1272 {\pm 7} )$ ~\\ $ f_{1}^{'}(1426.4 {\pm 71.30.9})$\\
\end{tabular} 
&\begin{tabular}{c}
~$1230$~\\ ~$1281.9$~\\ ~ $1273$ ~\\ ~$ 1512$~\\
\end{tabular}
 &\begin{tabular}{c}
~$980 $~\\ ~$980$~\\ ~ $1135 $ ~\\ ~$1315$~\\
\end{tabular}
&\begin{tabular}{c}$1171.78$~\\ ~$1173.78$~\\~ $1035.21$~\\~$1531.55$~\\ \end{tabular} 
  &\begin{tabular}{c}

\end{tabular} \\  
\hline
\end{tabular}
\caption{A comparison between the (pseudo)scalar and (axial)vector meson states as obtained in the present work (PLSM) and the corresponding results from PNJL \cite{P. Costa:PNJLA,P. Costa:PNJLB,P. Costa:PNJLC}, the latest compilation of PDG \cite{PDG:2018}, QMD/UrQMD \cite{URQM} and the lattice QCD calculations \cite{HotQCD,PACS-CS}.}
\label{masscomp}
\end{table} 
%}

\subsubsection{Dependence of various meson states on baryon chemical potential \label{sec:mudepnd}}

\begin{figure}[htb]
\centering{
\includegraphics[width=3.9cm,angle=-90]{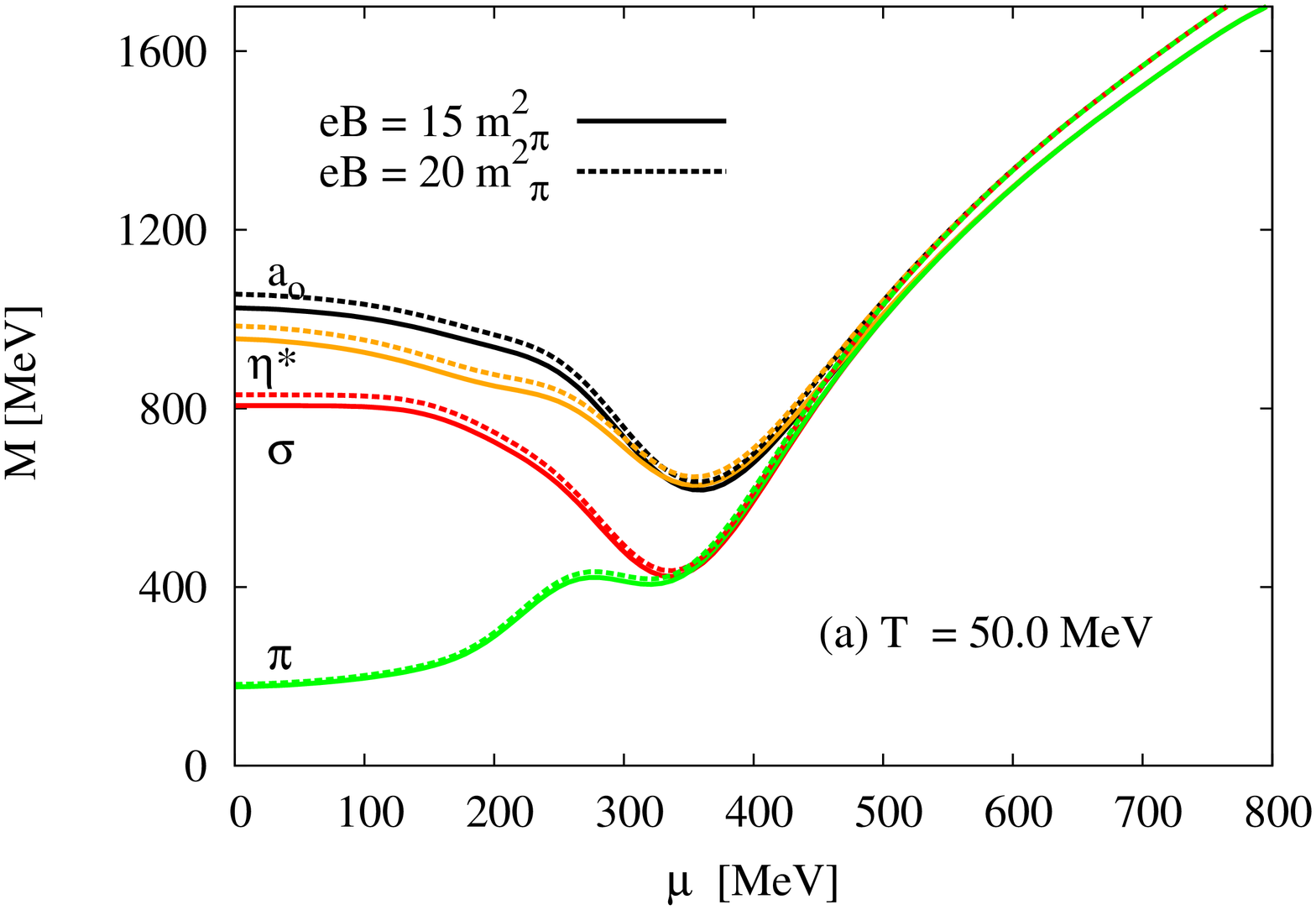}
\includegraphics[width=3.9cm,angle=-90]{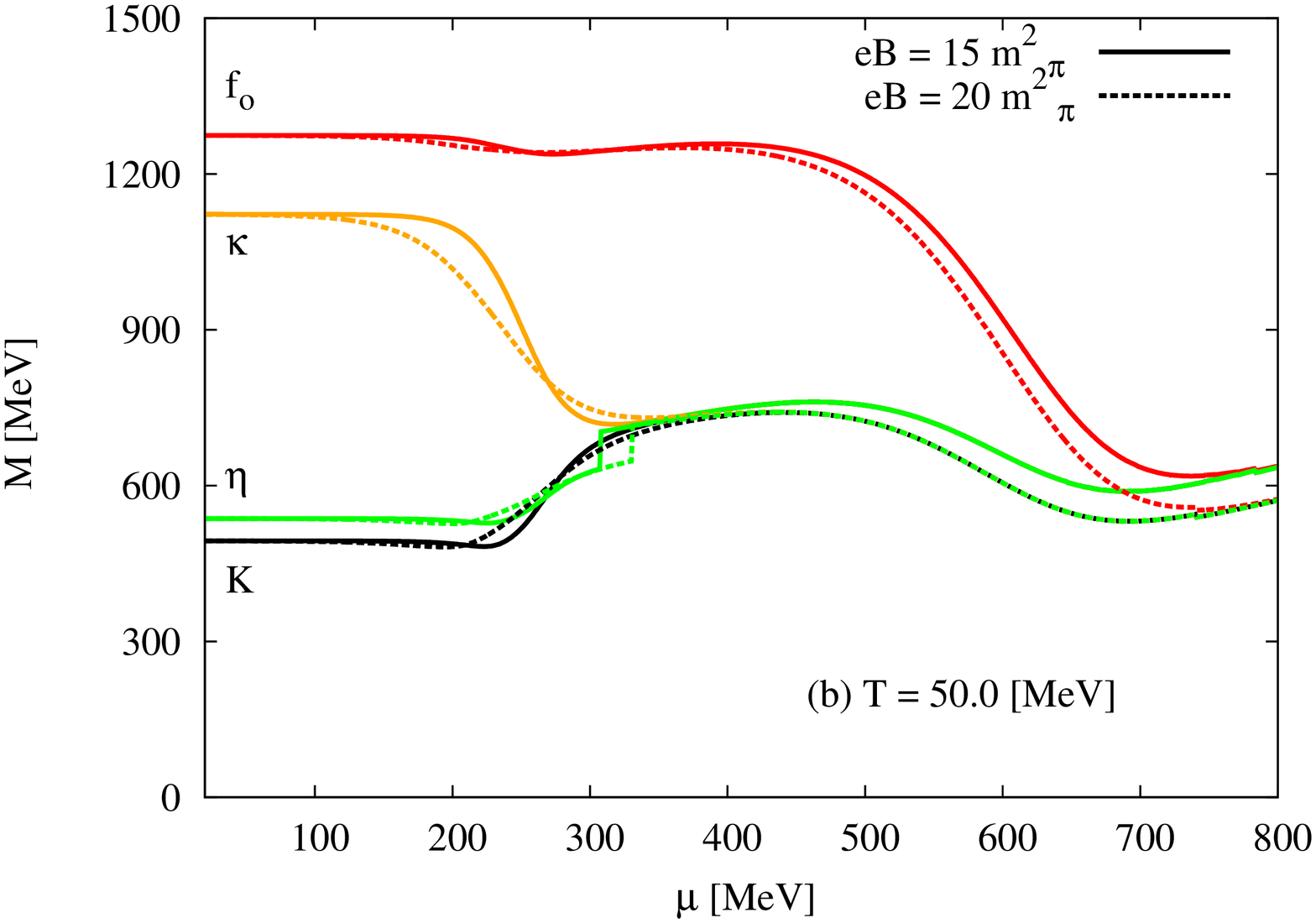}
\includegraphics[width=3.9cm,angle=-90]{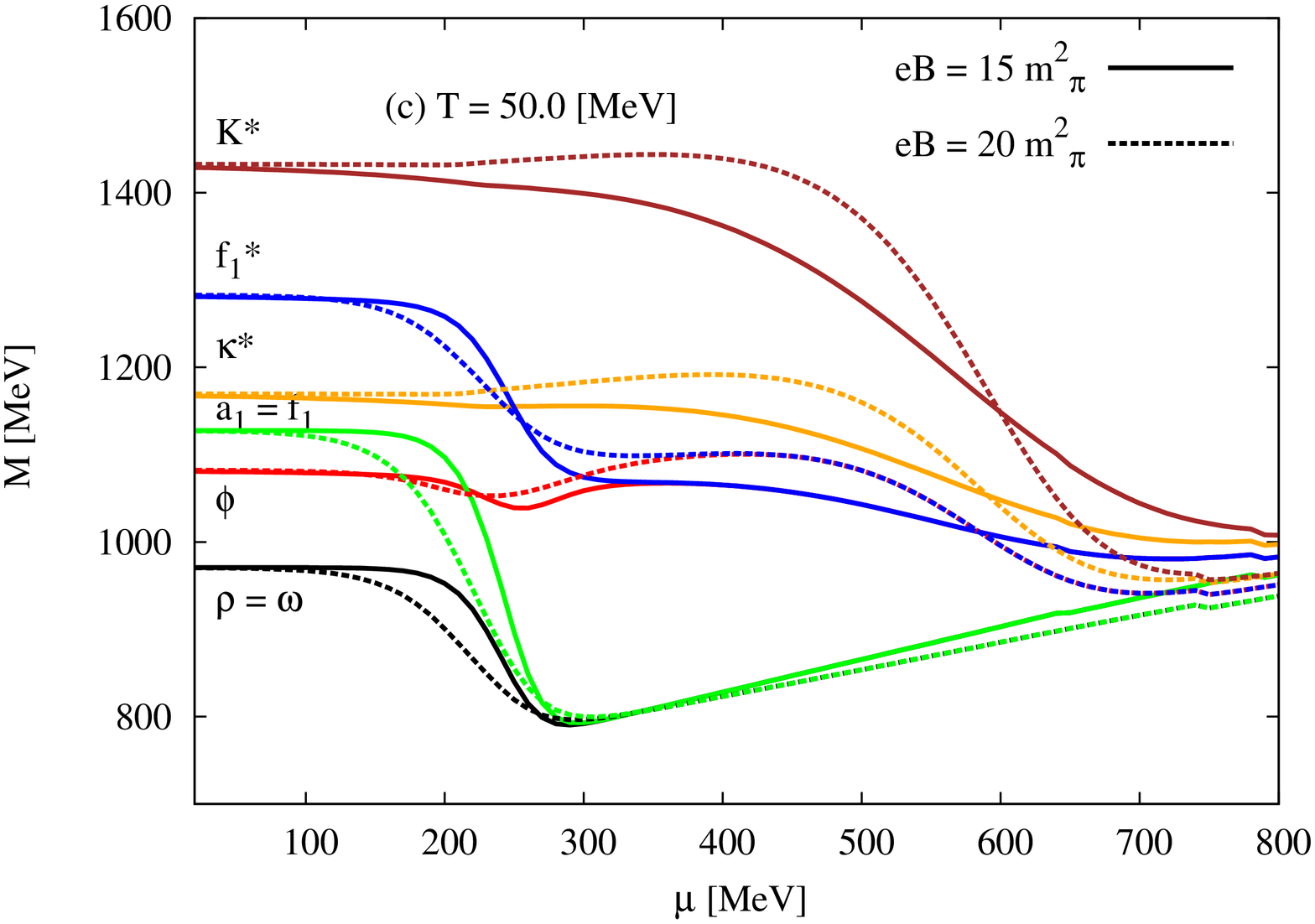}\\
\includegraphics[width=3.9cm,angle=-90]{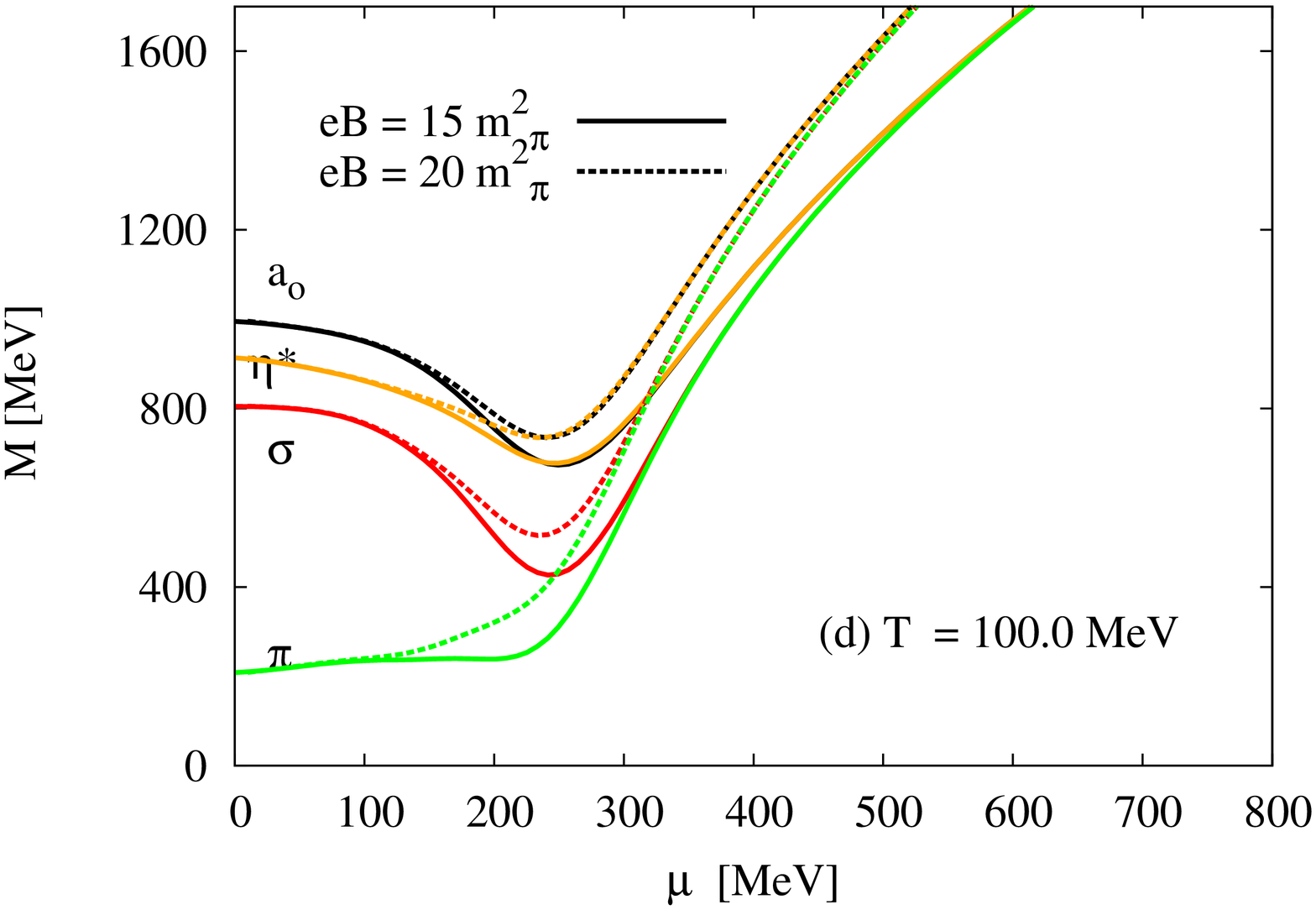}
\includegraphics[width=3.9cm,angle=-90]{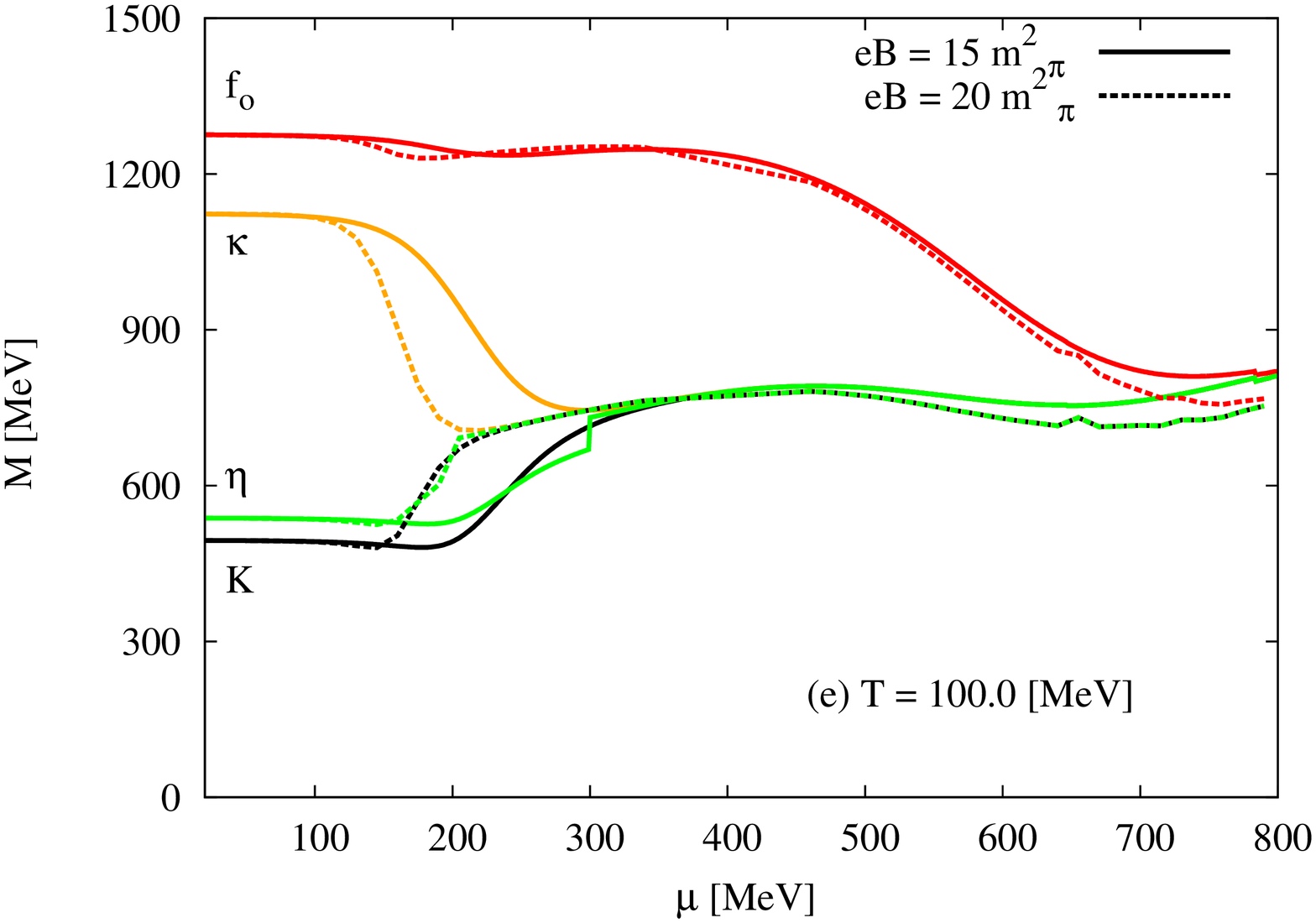}
\includegraphics[width=3.9cm,angle=-90]{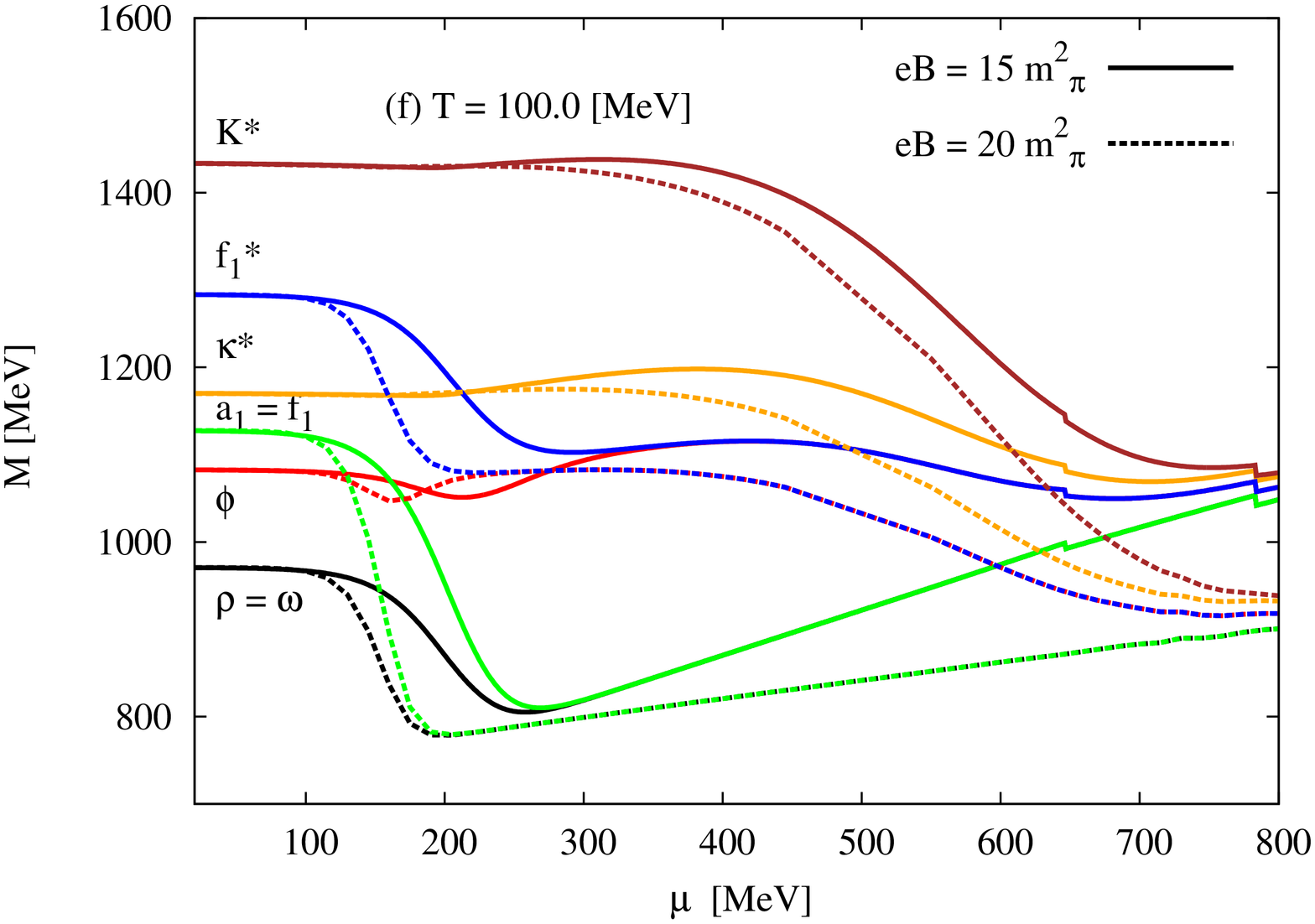}
\caption{\footnotesize (Color online) Masses of (pseduo)scalar and (axial)vector meson states are given in dependence on the baryon chemical potential at fixed magnetic field strengths; $eB=15\, m_{\pi}^2$ (solid curves) and $eB=20\, m_{\pi}^2$ (dotted curve) and at fixed temperatures;  $T=50~$MeV (top panels) and $T=100~$MeV (bottom panels). \label{fig:Mumass}
}}
\end{figure}

In presence of $U(1)_A$-anomaly, the dependence of the various meson states on the baryon chemical potential ($\mu$) at fixed temperatures; $T=50$ and $100~$MeV and fixed magnetic field strengths; $eB=15\, m_{\pi}^2$, and at $20\, m_{\pi}^2$ is depicted in Fig. \ref{fig:Mumass}. 

Figure \ref{fig:Mumass} shows the dependence of scalar and vector meson states (labeled curves) on the baryon chemical potentials, at fixed magnetic fields; $eB=15\, m_{\pi}$ (solid curves) and $eB=20\, m_{\pi}$ (dashed curves) and fixed temperatures; $T=50$ (top panel), and $100$MeV (bottom panel). As discussed in previous sections, the nonstrange  ($\sigma_l$) and strange quark ($\sigma_s$) chiral condensates strongly depend on the temperatures and on the baryon chemical potentials. The mass gap which distinguishes between the meson states is resulted from dense bosonic contributions. Such contributions remain independent on the baryon chemical potential until they reach the Fermi surface \cite{Tawfik:2014gga,Wagner:Thesis}. Then, the masses are liberated through a first-order phase transition. To assure that the Polyakov-loop potentials integrate deconfinement phase transition to LSM, the dynamics of color charges was integrated as a new type of interactions. Apparently, the meson masses remain degenerated through the deconfinement phase transition. When comparing these present results with our previous ones \cite{Tawfik:2014gga}, we conclude that the deconfinement phase transition in the present work, to which finite magnetic field is introduced, seems to become sharper and faster, i.e. the region of $\mu$ within which the masses of the meson states decrease indicating a rapid change in the underlying dynamics and the degrees of freedom, becomes narrower with introducing the magnetic field. This finding could be a subject of experimental verification, as well.

Another finding of the present work is that at a fixed magnetic field, the finite baryon chemical potential is conjectured to come up with {\it baryon} density contributions to the masses of various meson states. These are dominated in three regions:
\begin{itemize}
\item The first region are defined at low $\mu$, where the gap difference between the meson masses seems originated to the dense bosonic contributions. These are obtained from the mesonic Lagrangian of LSM.
\item The second region appears, when the meson states to feel/undergo chiral phase-transition from their confined hadron phase to parton phase, where the quarks and the gluons dominate the underlying degrees of freedom.
\item The third region is characterized by large $\mu$, where the hadrons are conjectured to dissolve into their constituents of free quarks and gluons.
\end{itemize}

As obtained while studying the temperature dependence, the dependence of the various meson masses on $\mu$ is accompanied by a decrease in the bosonic contributions to the medium density. These become negligible at very large baryon chemical potential. In this limit, the dense fermionic contributions become dominant. Furthermore, the latter come up with mass degeneration, which takes place through dense fermionic contributions. Accordingly, the meson states if they would pass through the phase transition as bound hadron states or survive the critical density ($\mu_{\mathrm{crit}}$), at which the phase transition takes place, remain in need of dense fermonic contributions in order to overcome the energy gap of Fermi surfaces. The latter differs from a meson state to another. With increasing $\mu$, the bound meson states are assumed to be liberated and entirely converted into deconfinment (partonic) phase \cite{Tawfik:2014gga,Wagner:Thesis}. 

In middle panels of Fig. \ref{fig:Mumass}, we notice that the meson masses drop through first-order phase transition. Concretely, the mass of $\mathrm{\kappa}$ state drops to the masses of $\mathrm{K}$ and that of $\mathrm{\eta}$ states at $\mu_{\mathrm{crit}} \sim 350~$MeV. The temperature is fixed to $50~$MeV. It seems that $\mathrm{f}_0$ meson state  has a stronger Fermi surface because of the strange quarks. The mass of this state seems to survive until $\mu \sim 500~$MeV. At $T=100$ (bottom panel), the first-order phase transition seems to last at small $\mu$. The mass of $\mathrm{\kappa}$ drops to the masses of $\mathrm{K}$ and that of $\mathrm{\eta}$ state at $\mu_{\mathrm{crit}} \sim 298~$MeV at magnetic field $eB=15\, m_{\pi}^2$ and $\mu_{\mathrm{crit}} \sim 200~$MeV at $eB=20\, m_{\pi}^2$. Also, the mass of $f_0$ meson degenerates at lower $\mu_{\mathrm{crit}}$ than that at $T=50~$MeV (top panel).

In Fig.  \ref{fig:Mumass}, two independent quantities ($T$ and $e\, B$) have non-negligible effects on the dependence of meson masses on $\mu$. From the QCD phase-diagram \cite{Tawfik:2016ihn,Tawfik:2017cdx} and the chemical freezeout boundary \cite{Tawfik:2014eba}, we conclude that increasing $T$ reduces $\mu_{\mathrm{crit}}$. Also, here we notice that increasing $T$ reduces $\mu_{\mathrm{crit}}$, at which the bosonic contribution at finite chemical potential becomes small. Increasing $\mu$ has an obvious influence on the scalar meson. This accelerates reaching stable levels of the meson masses. This is a level at which the meson masses no longer depend on $\mu$. Concretely, we observe that the masses of the vector mesons stay nearly constant until the phase transition takes place, while the masses of the axial vector mesons show a stronger melting above the Fermi surface. Furthermore, increasing $e B$ is assumed to reduce $\mu_{\mathrm{crit}}$ and thus enhance the chiral phase-transition. This explains the influences of finite $e B$ on the $\mu$-dependence of various meson states.

\begin{figure}[htb]
\centering{
\includegraphics[width=3.9cm,angle=-90]{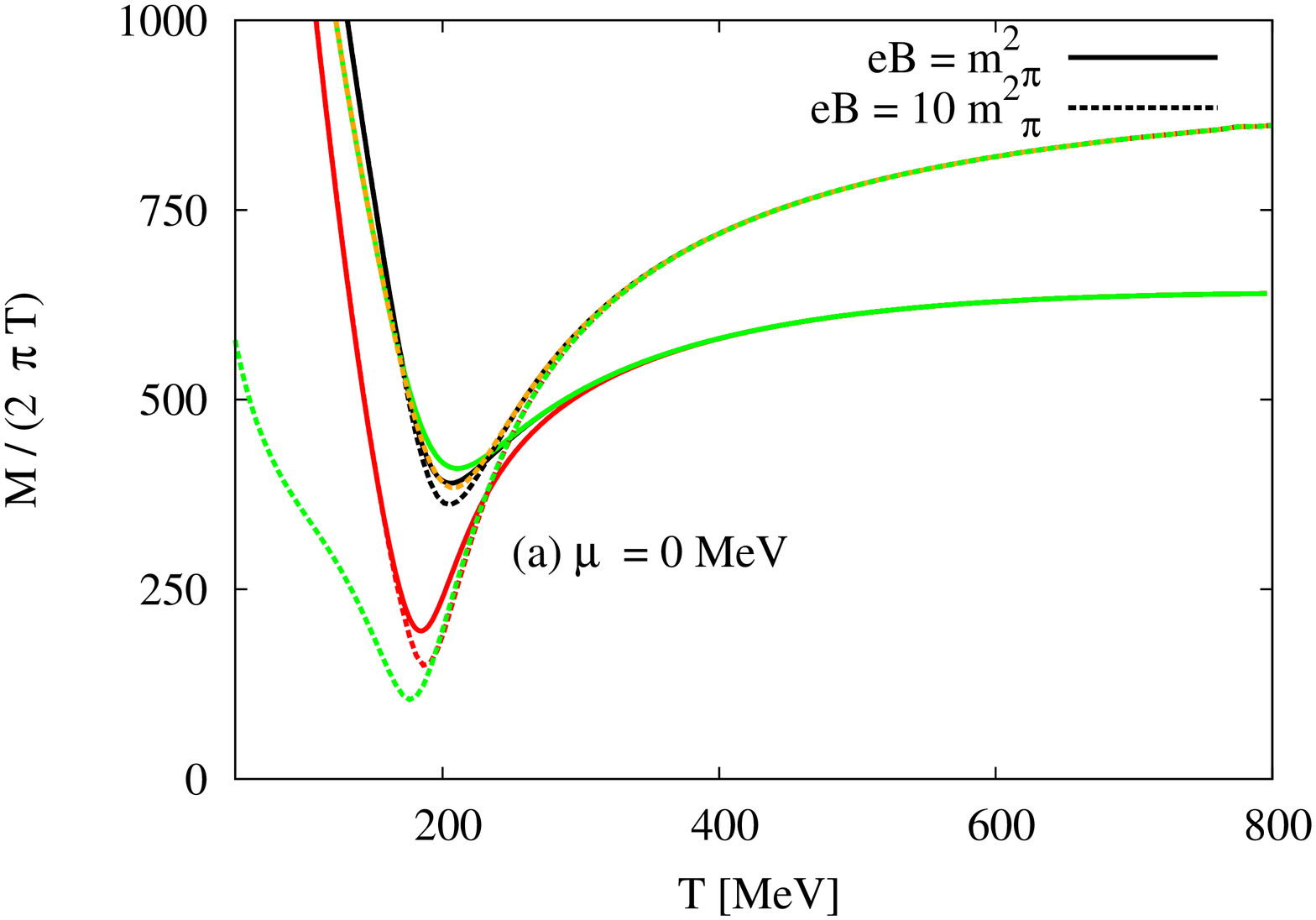}
\includegraphics[width=3.9cm,angle=-90]{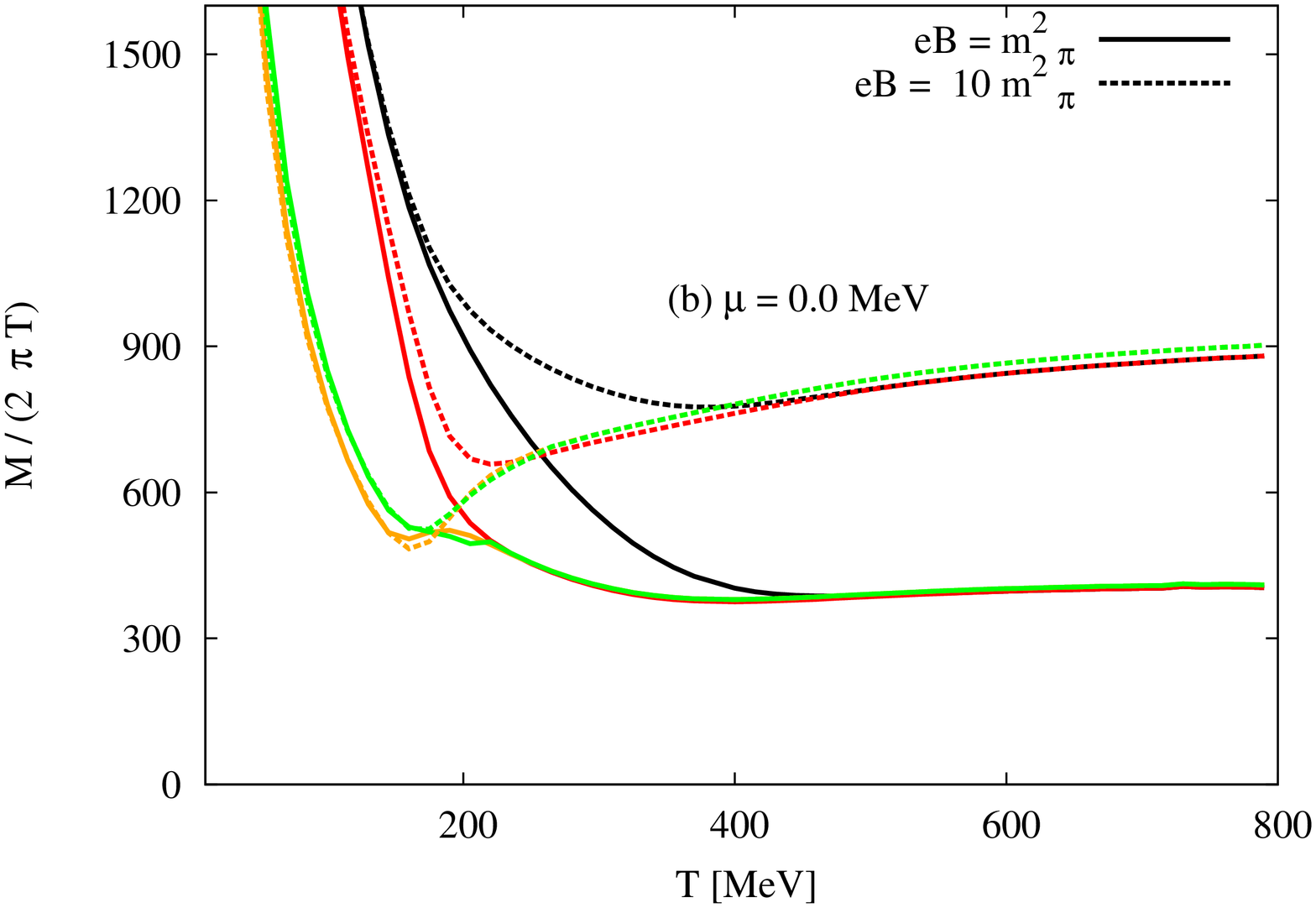}
\includegraphics[width=3.9cm,angle=-90]{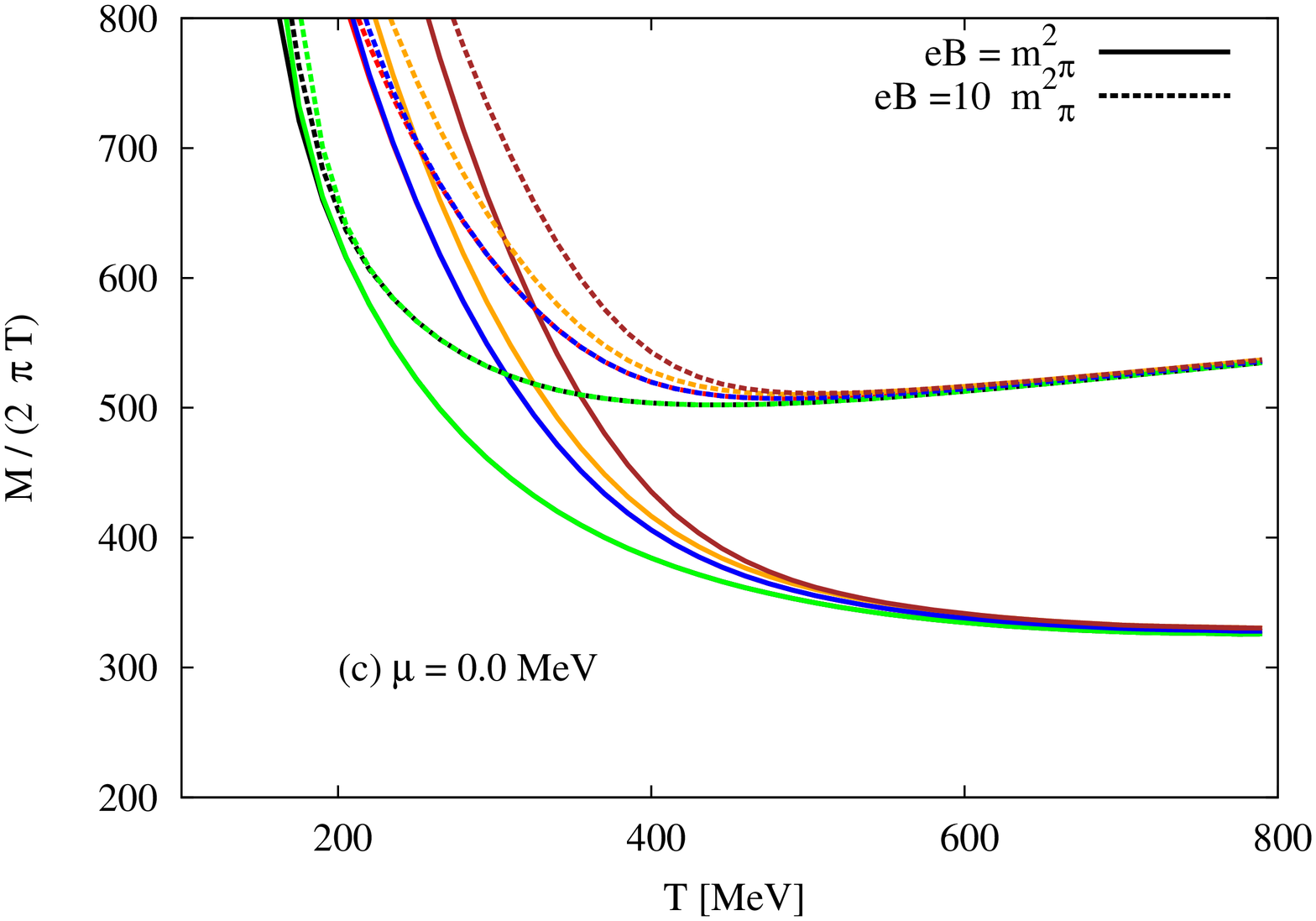}\\
\includegraphics[width=3.9cm,angle=-90]{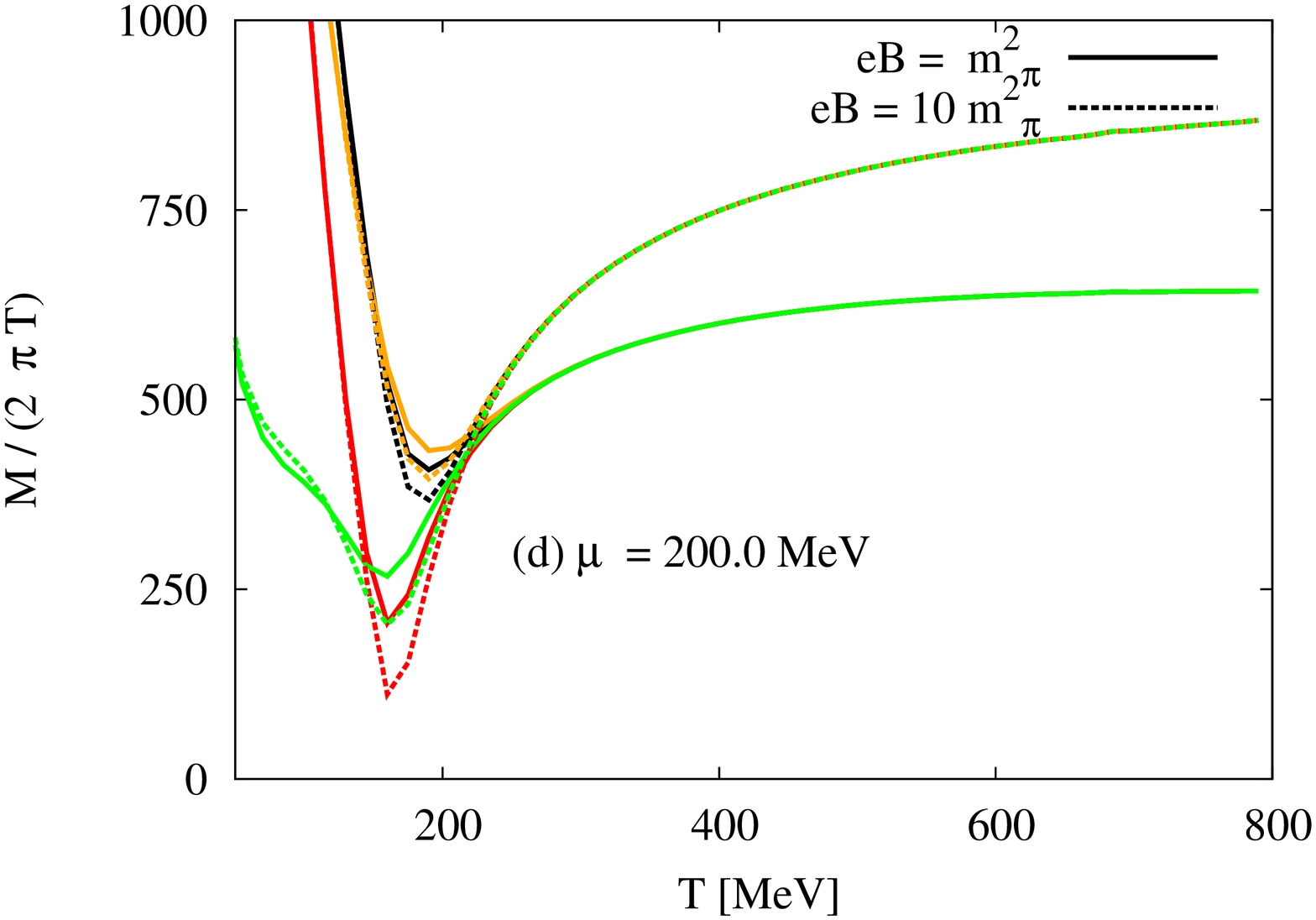}
\includegraphics[width=3.9cm,angle=-90]{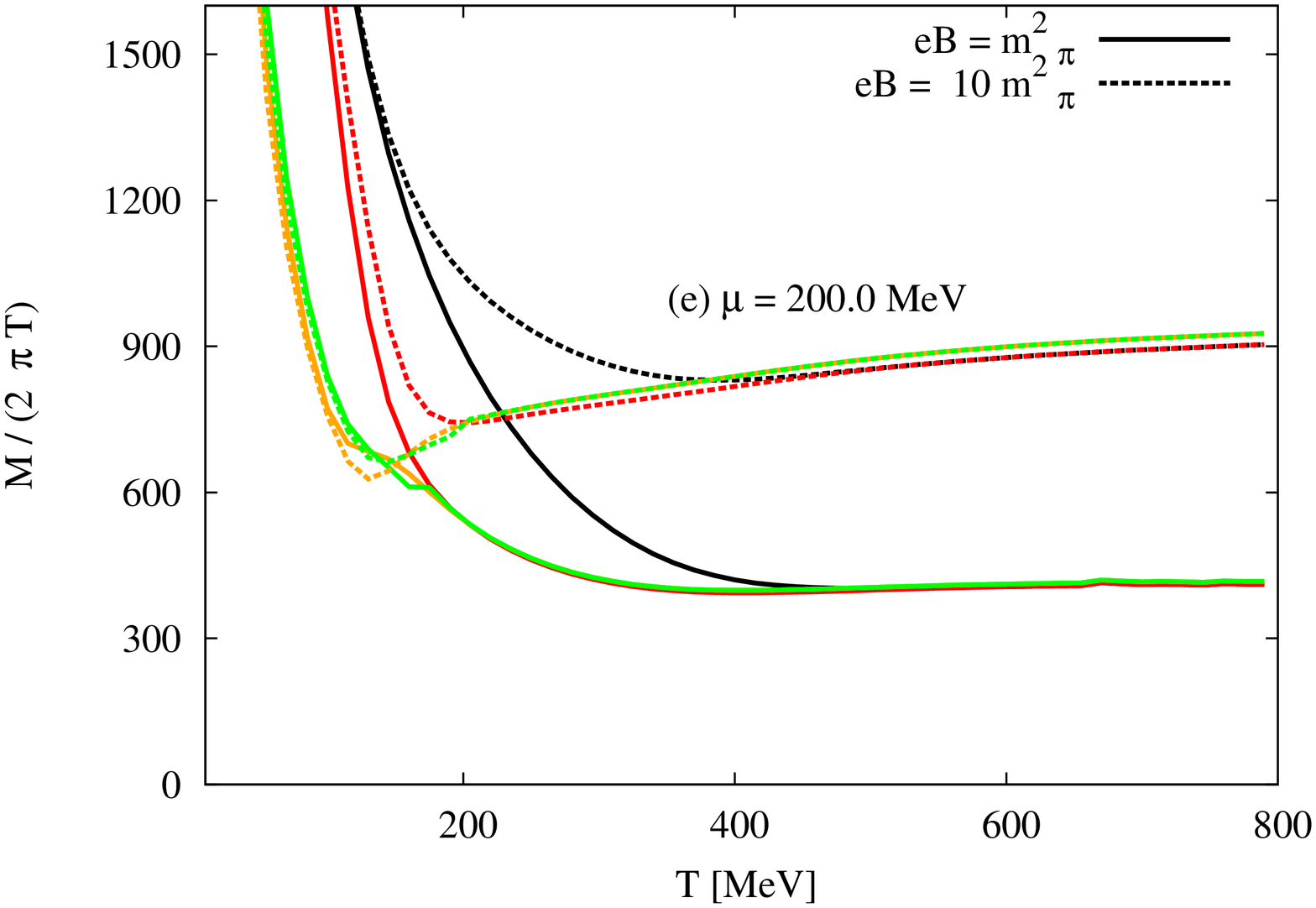}
\includegraphics[width=3.9cm,angle=-90]{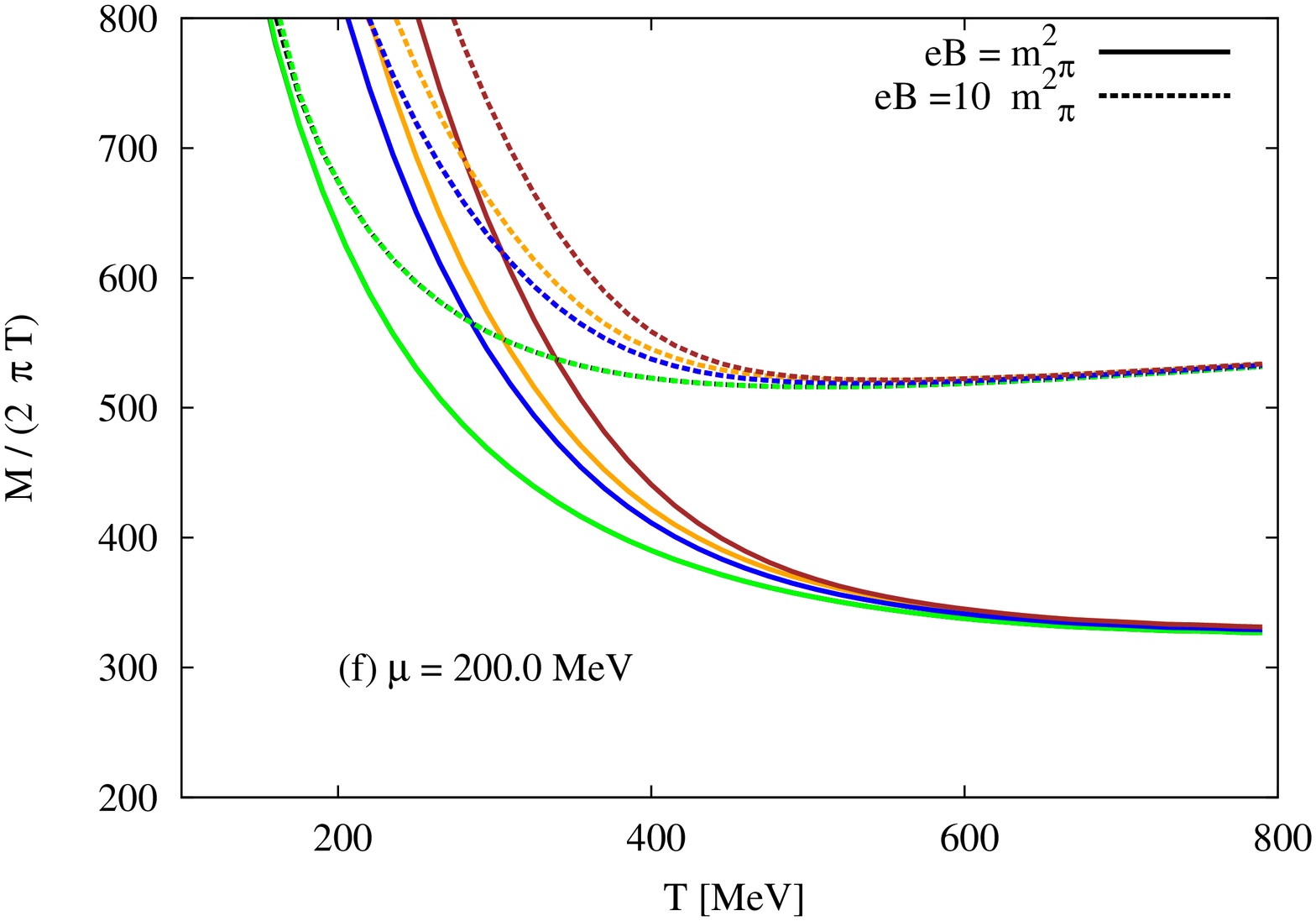}
\caption{\footnotesize (Color online)  Left-hand and middle panels depict the scalar and pseudoscalar meson sectors, respectively, as functions of temperatures at vanishing (top panels) and finite baryon chemical potentials (bottom panels). Right-hand panel shows the same as in left-hand panel but for vector and axialvector meson sectors. \label{matsubra}
}}
\end{figure}

\subsubsection{Various meson stated normalized to the lowest Matsubara frequencies \label{sec:lMatsubaraF}}

In thermal field theory, the Matsubara frequencies are considered as a summation over all discrete imaginary frequencies
\bea 
S_{\pm} &=& T \sum_{i \omega_n} g(i \omega_n), 
\eea
where $\pm$ stands for bosons and fermions, respectively, and $g(i \omega_n)$ is a rational function, with $\omega_n=2 n \pi T$ and $\omega_n=(2 n+1) \pi T$ represent bosons and fermions, respectively. The integers $n=0,\pm 1,\pm 2, \cdots$ play an important role on the quantization process. By using the Matsubara frequency, the weighting function $\Gamma_{\eta}(z)$ reaches two poles at $z=i \omega_n$. Then
\bea
S_{\pm} &=& \frac{T}{2 \pi i} \oint g(z) \, \Gamma_{\eta}(z) \, dz.
\eea
$\Gamma_{\pm}(z)$ can be chosen depending on which half plane the convergence is to be controlled in, 
\bea
\Gamma_{\pm}(z) &=& \le\{\begin{array}{l} \eta \, \frac{1+n_{\pm}(z)}{T} \\ \\ \eta\, \frac{n_{\pm}(z)}{T} \end{array} \ri.,
\eea
where the single particle distribution function is given as $n_{\pm}(z)=(1\pm e^{z/T})^{-1}$. 

The boson masses are conjectured to get contributions from the Matsubara frequencies \cite{lmf1}. Furthermore, at temperatures above the chiral phase-transition  ($T\geq T_c$), the thermal behavior of the thermodynamic quantities, such as susceptibilities, and that of the masses are affected by the interplay between the lowest Matsubara frequency and the Polyakov-loop potentials \cite{lmf2}. In light of this, we apply normalization to the present meson sectors with respect to lowest Matsubara frequency which is associated to bosons $2\, \pi\, T$ \cite{Tawfik:2006by}. This would allow us to define the dissolving {\it critical} temperatures. At this temperature, the various meson states are conjectured to liberate into degrees of freedom of free quarks and free gluons. We found that the different meson states have different dissolving temperatures. This means that the different hadrons, in this case mesons, have different chiral critical temperatures ($T_{\chi}$). In other words, not all hadrons {\it melt} into quark-gluon plasma, simultaneously.

Figure \ref{matsubra} depicts the normalization of the different meson states with respect to the lowest - in this case {\it bosonic} - Matsubara frequencies ($2\, \pi\, T$) at vanishing (top panel) and finite baryon chemical potential (bottom panel) in different magnetic fields; $eB=m_{\pi}^2$ (solid curves) and $eB=10\, m_{\pi}^2$ (dotted curves). Here, the legends are identical to the ones shown in previous figures. It is obvious that the masses of almost all meson states become independent on temperature, i.e. constructing a kind of a universal curve, especially at high temperatures. This observation gives a clear signature for the dissociation of mesons which are apparently dissolving into {\it free} quarks and gluons. Also, it is found that the characteristic temperature, the critical temperature, seems not being universally valid for all meson states. As discussed in earlier sections, there are two different thermodynamics variables; the temperatures and the magnetic fields, seem to enhance the chiral phase-transition and therefore come up with a considerable influence on the corresponding critical temperature. This observation is also confirmed here. 

Table \ref{dissolve} summarizes the different  meson states corresponding and their approximate dissolving {\it critical} temperatures ($T_c^d$) in MeV-units in finite magnetic field $eB=m_{\pi}^2$ but at vanishing baryon chemical potential. A magnetic field strengths $eB=m_{\pi}^2$ can be likely generated at RHIC top energies. The critical temperatures $T_c^d$ can roughly be read off from the resulting graph. Thus, the statistical uncertainties (errors) are not estimated, yet. As mentioned, the different meson states seem to have different dissolving temperatures. This is a novel observation of the present work. To connect the latter with the critical temperatures, the first principle lattice calculations should be first reproduced by our QCD-like calculations. The experimental results should be verified, as well. But, it should be noticed that, our calculations are based on the QCD-like model, the PLSM, in which not all aspects of QCD are taken into account. Nevertheless, PLSM - as such - seems to draw a good picture about what the lattice QCD simulations produce. As an outlook, we plan to devote a future work to a systematic analysis for the resulting dissolving temperatures at varying magnetic fields. The latter can be achieved at different beam energies and in different heavy-ion collisions centralities, especially by means of the future facilities which shall enable us covering the gap at high densities.

%%%%%%%%%%%%%%%%%%%%%%%%%%%%%%%%%%%%%%%%%%%%%%%
%%%   Table :- Disolving  hadrons 
%%%%%%%%%%%%%%%%%%%%%%%%%%%%%%%%%%%%%%%%%%%%%%%
\begin{table}[htb]
\begin{center}
\begin{tabular}{ ||c | c | c | c | c|| }
\hline 
Comparison  & Scalar Mesons & Pseudoscalar  Mesons & Vector  Mesons & Axialvector Mesons  \\ 
\hline  \hline
Meson States 
&
\begin{tabular}{cccc}
$a_{0}$~& ~$\kappa $~&~$\sigma$~&~$ f_{0}$~\\
\end{tabular}
 & 
\begin{tabular}{cccc}
~$\pi$~&~ $K $~ &~ $\eta$ ~& ~$\eta ^{'}$~\\
\end{tabular} 
 &
 \begin{tabular}{cccc}
~$\rho$~& ~$K _{0}^{*}$~ & ~$\omega$~ &~ $ \phi$~\\
\end{tabular}
 & 
\begin{tabular}{cccc}
~$a_{1}$~ &~ $K_1 $ ~& ~$f_1$~& ~ $ f_1^*$~\\ 
\end{tabular} \\ 
\hline 
\begin{tabular}{cccc}
$T_{c}^{d}$ in MeV 
\end{tabular} 
&
\begin{tabular}{c c c c}
$ 430$ & $450$ & $470$ & $450$ \\
\end{tabular}
 & 
\begin{tabular}{c c c c}
$ 320$ &$ 230$ & $ 335$ &  $ 240$ \\
\end{tabular} 
 &
\begin{tabular}{c c c c}
$ 495$ &$ 495$ & $ 495$& $   495$\\
\end{tabular}
 & 
\begin{tabular}{c c c c}
$ 495$ &$ 495$ & $ 495$& $   495$\\ 
\end{tabular} \\ \hline 
\end{tabular}
\caption{Approximate dissolving {\it critical} temperatures, $T_c^d$, corresponding to the different meson states at $eB=m_\pi^2$. The statistical uncertainties (errors) are not estimated, yet. \label{dissolve}}
\end{center}
\end{table}

%%%%%%%%%%%%%%%%%%%%%%%%%%%%%%%%%%%%%%%%%%%%%
\section{Conclusion \label{conclusion}}
%%%%%%%%%%%%%%%%%%%%%%%%%%%%%%%%%%%%%%%%%%%%%

The present study aims at presenting a systematic investigation for the temperature and the chemical potential dependence of sixteen meson states [pseudoscalars ($J^{pc}=0^{-+}$), scalars ($J^{pc}=0^{++}$), vectors ($J^{pc}=1^{--}$) and axial-vectors ($J^{pc}=1^{++}$)] constructed from the SU($3$) Polyakov linear-sigma model in presence of finite magnetic fields. The introduction of magnetic fields to this QCD-like model is accompanied by certain modifications in the phase space and in the dispersion relation, among others. Also, this requires Landau quantization to be implemented, for which we use Landau theory for quantized cyclotron orbits of charged particles in external magnetic-field. Consequently, some restrictions are added to the color and electric charges of the quarks. 

In relativistic heavy-ion collisions, a huge magnetic field is very likely expected. For instance, due to oppositely directed relativistic motion of charges (spectators) in peripheral collisions and/or due to the local momentum-imbalance of the participants in central collisions, a huge magnetic field can be created. To have a picture about its magnitude, we recall that at SPS, RHIC and LHC energies, the strength of such fields ranges between $0.1$ to $1$ to $10-15 \, m_{\pi}^2$, respectively, where $m_{\pi}^2 \sim 10^8~$Gauss. In light of this, studying its possible influences on the QCD matter is of great impacts on our understanding of the heavy-ion collisions and therefore increasingly gains popularity among particle physicists.

In mean-field approximation of PLSM, a grand canonical partition function can be constructed. In order to prepare for the proposed calculations, we have to estimate the temperature dependence of the deconfinement ($\phi$ and $\phi^*$) and chiral order parameters ($\sigma_l$ and $\sigma_s$) in presence of finite magnetic field. We have found that, taking into consideration finite magnetic field influences all these parameters and according;y strongly affect the entire QCD  phase transition.  
 
In a previous work, we have studied the distribution of Landau levels and shown how they are occupied at finite magnetic field, temperature, and baryon chemical potential \cite{Tawfik:2016ihn,Tawfik:2017cdx}. We have concluded that the occupation of each Landau level varies with the quark electric charge besides the temperature and the baryon chemical potential. This is assumed to be characterized by QCD energy scale.

In PLSM with mean-field approximation, the masses can be determined form the second derivative of the free energy with respect to the corresponding hadron field of interest evaluated at its global minima. We have presented temperature and chemical potential dependence of SU($3$) sixteen meson masses, at finite magnetic fields. At finite magnetic field, the broken chiral-symmetry is assumed to contribute to the mass spectra. We utilize this in defining the chiral phase structure of nonent meson sectors as functions of temperatures and chemical potentials. We found that the spectra of the meson masses can be divided into three regions. \begin{itemize}
\item At small temperatures and baryon chemical potentials, the first region is related to the bosonic contributions which are apparently stable. 
\item The second region is the chiral phase transition, which seems to have a remarkable influence stemming from the finite magnetic field. It is obvious that, the finite magnetic field improves, i.e. sharpens and accelerates, the chiral phase transition of the given meson state. 
\item The third region is characterized by fermionic contributions which increase with increasing the temperature.
\end{itemize}

We conclude that the thermal bosonic (meson) mass contributions decreases with increasing temperature. The fermionic (quark) contributions considerably increases at high temperature.  At low temperatures, the bosonic contributions become dominant and reach a kind of {\it stable} plateaus relative to the vacuum mass of each meson sector. In this temperature-limit, the effect of the fermionic contributions are negligibly small. It noteworthy highlighting that the bosonic contributions result in the mass gap between the different meson states in thermal and dense medium. With increasing temperature, the fermionic contributions complete the thermal behavior of these states and lead to mass degeneracy at very high temperatures. 

From scalar and vector mesons normalized to the lowest Matsubara frequency, we also conclude that a rapidly decrease in their masses is observed as the temperature increases. Then, starting from the critical temperature corresponding to each meson sector, we find that the temperature dependence almost vanishes. At high temperatures, we notice that, the masses of almost all meson states become temperature independent, i.e. they are constructing a kind of a universal line. This is to be seen as a signature for dissolving the confined mesons into colored quarks and gluons. In other words, the meson states undergo different deconfimement phase transitions, i.e. the various hadrons likely have various {\it critical} temperatures. This is one of the essential findings of the present study to be confirmed by the first-principle lattice simulations and ultimately in future experiments. 

We have  compared our calculations on scalar and vector mesons with the latest compilation of PDG, the lattice QCD calculations, and QMD/UrQMD simulations and found that our results are remarkably precise, especially for some light mesons at vanishing temperature. This implies that the parameters of the QCD-like model we have utilized are reliable.

\appendix
\section{Masses of sixteen mesonic-states \label{app:Mass}}
\label{Meson}

In vacuum, the mesonic sectors are formulated in dependence on the nonstrange and strange fields:
\begin{itemize}
\label{massesALL}
\item{Scalar meson states}, the squared masses for the scalar sector ($i=S$) are given as  
\bea
  m^2_{a_0} &=& m^2 + \lambda_1 \le(\bsig_l^2 + \bsig_s^2\ri) + \frac{3
    \lambda_2}{2} \bsig_l^2 +\frac{\sqrt{2} c}{2} \bsig_s ,\\
  m^2_{\kappa} &=& m^2 + \lambda_1 \le(\bsig_l^2 + \bsig_s^2\ri) +
  \frac{\lambda_2}{2} \le(\bsig_l^2 + \sqrt{2} \bsig_l \bsig_s +
    2 \bsig_s^2\ri) + \frac{c}{2}\bsig_l, \label{mkappa}\\
  m^2_\sigma &= &m^2_{S,{00}} \cos^{2}\theta _S +
  m^2_{S,{88}}\sin^{2}\theta_S + 2  m^2_{S,{08}} \sin \theta_S \cos
  \theta_S ,\\
  m^2_{f_0}  &=& m^2_{S,{00}} \sin^{2}\theta _S +
  m^2_{S,{88}}\cos^{2}\theta_S - 2  m^2_{S,{08}} \sin \theta_S \cos \theta_S , 
\eea
with 
\bea 
m^2_{S,{00}} &=&m^2 + \frac{\lambda_1}{3} \le(7 \bsig_l^2 + 4
  \sqrt{2}\bsig_{l}\bsig_{s} + 5 \bsig_s^2\ri) + \lambda_2\le(\bsig_l^2 +
  \bsig_s^2\ri) -\frac{\sqrt{2}c}{3}\le(\sqrt{2} \bsig_l +
  \bsig_s\ri),\nonumber \\
 m^2_{S,{88}} &=& m^2 + \frac{\lambda_1}{3} \le(5 \bsig_l^2 - 4
  \sqrt{2}\bsig_{l}\bsig_{s} + 7 \bsig_s^2\ri) +
  \lambda_2\le(\frac{\bsig_l^2}{2} + 2\bsig_s^2\ri)
  +\frac{\sqrt{2}c}{3}\left(\sqrt{2} \bsig_l -
  \frac{\bsig_s}{2}\right),\nonumber  \\
  m^2_{S,{08}} &=& \frac{2 \lambda_1}{3}\le(\sqrt{2}\bsig_l^2 - \bsig_l \bsig_s - \sqrt{2}\bsig_s^2 \ri)
  + \sqrt{2}\lambda_2 \le(\frac{\bsig_l^2}{2} - \bsig_s^2 \ri)
  + \frac{c}{3\sqrt{2}} \le(\bsig_l - \sqrt{2}\bsig_s\ri). \nn
\eea
\item{Pseudoscalar meson states}, the squared masses for the pseudoscalar sector ($i=p$) read
\bea
  m^2_\pi &=& m^2 + \lambda_1 \le(\bsig_l^2 + \bsig_s^2\ri) +
  \frac{\lambda_2}{2} \bsig_l^2 -\frac{\sqrt{2} c}{2} \bsig_s, \\
  m^2_K &=& m^2 + \lambda_1 \le(\bsig_l^2 + \bsig_s^2\ri) +
  \frac{\lambda_2}{2} \le(\bsig_l^2 - \sqrt{2} \bsig_l \bsig_s +
    2 \bsig_s^2\ri) - \frac{c}{2} \bsig_l, \\
    m^2_{\eta'} &=& m^2_{p,{00}} \cos^{2}\theta _p + m^2_{p,{88}}\sin^{2}\theta _p + 2 m^2_{p,{08}} \sin \theta _p \cos \theta_p, \\
  m^2_{\eta} &=& m^2_{p,{00}} \sin^{2}\theta_p + m^2_{p,{88}}\cos^{2}\theta_p - 2 m^2_{p,{08}} \sin \theta_p \cos \theta_p,
\eea
with
\bea
m^2_{p,{00}} &=& m^2 + \lambda_1 \le(\bsig_l^2 + \bsig_s^2\ri)
  + \frac{\lambda_2}{3}\le(\bsig_l^2 + \bsig_s^2\ri) +\frac{c}{3}\le(2
  \bsig_l + \sqrt{2} \bsig_s\ri), \nonumber \\
  m^2_{p,{88}} &=& m^2 + \lambda_1 \le(\bsig_l^2 + \bsig_s^2\ri)
  + \frac{\lambda_2}{6}\le(\bsig_l^2 + 4 \bsig_s^2\ri) -\frac{c}{6}\le(4
  \bsig_l - \sqrt{2}\bsig_s\ri), \nonumber \\
  m^2_{p,{08}} &=& \frac{\sqrt{2} \lambda_2}{6} \le(\bsig_l^2 - 2
  \bsig_s^2\ri) - \frac{c}{6}\le(\sqrt{2}\bsig_l - 2 \bsig_s\ri), \nonumber
\eea
and the mixing angles are given by
\bea
  \tan 2\theta_i &=& \frac{2 m^2_{i,{08}}}{m^2_{i,{00}}-m^2_{i,{88}}}\
  ,\  i=S,p\ .
\eea
\item{The vector meson states}, the squared masses for the vector sector ($i=V^{\mu}$) can be expressed as
\bea
m_{\rho}^{2}  &=&m_{1}^{2}+\frac{1}{2} \le(h_{1}+h_{2}+h_{3}\ri) \bsig_l^{2} 
+\frac{h_{1}}{2}\bsig_s^{2}+2\delta_{l}\;,\label{m_rho}\\
m_{K^{\star}}^{2}  &=&m_{1}^{2}+\frac{\bsig_l^{2}}{4} \le(g_{1}^{2}+2h_{1}
+h_{2}\ri) +\frac{\bsig_l \bsig_s}{\sqrt{2}}(h_{3}-g_{1}^{2})+\frac{\bsig_s^{2}}{2}\le(g_{1}^{2}+h_{1}+h_{2}\ri)+\delta_{l}+\delta_{s}\;,\label{m_K_star}\\
m_{\omega_{x}}^{2}  &=&m_{\rho}^{2}\;,\\
m_{\omega_{y}}^{2}  &=&m_{1}^{2}+\frac{h_{1}}{2}\bsig_l^{2} + \le(
\frac{h_{1}}{2}+h_{2}+h_{3}\ri) \bsig_s^{2}+2\delta_{s}\;,
\label{V}%
\eea
\item{And finally the axial--vector meson states}, the squared masses for the axial-vector sector ($i=A^{\mu}$) are
\bea
m_{a_{1}}^{2}  &=&m_{1}^{2}+\frac{1}{2} \le(2g_{1}^{2}+h_{1}+h_{2}-h_{3}\ri) \bsig_l^{2}+\frac{h_{1}}{2}\bsig_s^{2}+2\delta_{l} ,\label{m_a_1}\\
m_{K_{1}}^{2}  &=&m_{1}^{2}+\frac{1}{4}\le(  g_{1}^{2}+2h_{1}+h_{2}\ri)
\bsig_l^{2}-\frac{1}{\sqrt{2}}\bsig_l \bsig_s \le(h_{3}-g_{1}^{2}\ri) + \frac{1}{2}\le(g_{1}^{2}+h_{1}+h_{2}\ri)  \bsig_s^{2} \nn \\ 
&+&\delta_{l}+\delta_{s} ,\label{m_K_1}\\
m_{f_{1x}}^{2}  &=& m_{a_{1}}^{2}, \\
m_{f_{1y}}^{2}  &=&m_{1}^{2}+\frac{\bsig_l^{2}}{2} h_{1}+\le(2g_{1}^{2}+\frac{h_{1}}{2}+h_{2}-h_{3}\ri)  \bsig_s^{2}+2\delta_{s}.
\label{AV}%
\eea
\end{itemize}

%%%%%%%%%%%%%%%%%%%%%%%%%%%%%%%%%%%%%%%%%%%%%
%%%             references
%%%%%%%%%%%%%%%%%%%%%%%%%%%%%%%%%%%%%%%%%%%%%

\end{document}